\documentclass[12pt]{article}
\begin{document}
\newcommand{\beq}{\begin{equation}}
\newcommand{\eeq}{\end{equation}}
\newcommand{\beqa}{\begin{eqnarray}}
\newcommand{\eeqa}{\end{eqnarray}}
\newcommand{\beqar}{\begin{eqnarray*}}
\newcommand{\eeqar}{\end{eqnarray*}}
\newcommand{\ben}{\begin{enumerate}}
\newcommand{\een}{\end{enumerate}}
\newcommand{\bit}{\begin{itemize}}
\newcommand{\eit}{\end{itemize}}
\newcommand{\al}{\alpha}
\newcommand{\be}{\beta}
\newcommand{\del}{\delta}
\newcommand{\D}{\Delta}
\newcommand{\eps}{\epsilon}
\newcommand{\ga}{\gamma}
\newcommand{\Ga}{\Gamma}
\newcommand{\ka}{\kappa}
\newcommand{\inn}{\!\cdot\!}
\newcommand{\h}{\eta}
\newcommand{\kk}{\varphi}
\newcommand\F{{}_3F_2}
\newcommand{\la}{\lambda}
\newcommand{\La}{\Lambda}
\newcommand{\na}{\nabla}
\newcommand{\Om}{\Omega}
\newcommand{\p}{\phi}
\newcommand{\sig}{\sigma}
\renewcommand{\t}{\theta}
\newcommand{\z}{\zeta}
\newcommand{\ssc}{\scriptscriptstyle}
\newcommand{\eg}{{\it e.g.,}\ }
\newcommand{\ie}{{\it i.e.,}\ }
\newcommand{\labell}[1]{\label{#1}} 
\newcommand{\reef}[1]{(\ref{#1})}
\newcommand\prt{\partial}
\newcommand\veps{\varepsilon}
\newcommand\ls{\ell_s}
\newcommand\cF{{\cal F}}
\newcommand\cA{{\cal A}}
\newcommand\cS{{\cal S}}
\newcommand\cT{{\cal T}}
\newcommand\cC{{\cal C}}
\newcommand\cL{{\cal L}}
\newcommand\cM{{\cal M}}
\newcommand\cN{{\cal N}}
\newcommand\cG{{\cal G}}
\newcommand\cI{{\cal I}}
\newcommand\cJ{{\cal J}}
\newcommand\cl{{\iota}}
\newcommand\cP{{\cal P}}
\newcommand\cQ{{\cal Q}}
\newcommand\cg{{\it g}}
\newcommand\cR{{\cal R}}
\newcommand\cB{{\cal B}}
\newcommand\cO{{\cal O}}
\newcommand\cK{{\cal K}}
\newcommand\cH{{\cal H}}
\newcommand\tcO{{\tilde {{\cal O}}}}
\newcommand\bz{\bar{z}}
\newcommand\bw{\bar{w}}
\newcommand\hF{\hat{F}}
\newcommand\hA{\hat{A}}
\newcommand\hT{\hat{T}}
\newcommand\htau{\hat{\tau}}
\newcommand\hD{\hat{D}}
\newcommand\hf{\hat{f}}
\newcommand\hg{\hat{g}}
\newcommand\hp{\hat{\phi}}
\newcommand\hi{\hat{i}}
\newcommand\ha{\hat{a}}
\newcommand\hQ{\hat{Q}}
\newcommand\hP{\hat{\Phi}}
\newcommand\hS{\hat{S}}
\newcommand\hX{\hat{X}}
\newcommand\tL{\tilde{\cal L}}
\newcommand\hL{\hat{\cal L}}
\newcommand\tG{{\widetilde G}}
\newcommand\tg{{\widetilde g}}
\newcommand\tphi{{\widetilde \phi}}
\newcommand\tPhi{{\widetilde \Phi}}
\newcommand\te{{\tilde e}}
\newcommand\tk{{\tilde k}}
\newcommand\tf{{\tilde f}}
\newcommand\tF{{\widetilde F}}
\newcommand\tK{{\widetilde K}}
\newcommand\tE{{\widetilde E}}
\newcommand\tpsi{{\tilde \psi}}
\newcommand\tX{{\widetilde X}}
\newcommand\tD{{\widetilde D}}
\newcommand\tO{{\widetilde O}}
\newcommand\tS{{\tilde S}}
\newcommand\tB{{\widetilde B}}
\newcommand\tA{{\widetilde A}}
\newcommand\tT{{\widetilde T}}
\newcommand\tC{{\widetilde C}}
\newcommand\tV{{\widetilde V}}
\newcommand\thF{{\widetilde {\hat {F}}}}
\newcommand\Tr{{\rm Tr}}
\newcommand\tr{{\rm tr}}
\newcommand\STr{{\rm STr}}
\newcommand\M[2]{M^{#1}{}_{#2}}
\newcommand\nn{\nonumber}
\newcommand\lan{\langle}
\newcommand\ran{\rangle}
\parskip 0.3cm

\vspace*{1cm}

\begin{center}
{\bf \Large     Sphere-level Ramond-Ramond  couplings   \\ in Ramond-Neveu-Schwarz formalism }

\vspace*{1cm}

{Hamid R. Bakhtiarizadeh\footnote{hamidreza.bakhtiarizadeh@stu-mail.um.ac.ir} and {Mohammad R. Garousi\footnote{garousi@um.ac.ir}}}\\
\vspace*{1cm}
{ \it Department of Physics, Ferdowsi University of Mashhad,\\ P.O. Box 1436, Mashhad, Iran }

\vskip 0.6 cm

\vspace{2cm}
\end{center}

\begin{abstract}
\baselineskip=18pt
We calculate in details the  sphere-level scattering amplitude of two Ramond-Ramond (RR) and two Neveu-Schwarz-Neveu-Schwarz (NSNS) vertex operators in type II superstring theories   in Ramond-Neveu-Schwarz (RNS) formalism. We then compare   the expansion of this  amplitude  at order $\alpha'^3$   with the eight-derivative couplings of the gravity and B-field   that have been recently found based on  S-dual and T-dual Ward identities. We find exact agreement. Moreover, using the above S-matrix element, we have     found   various couplings involving the dilaton field, and shown that they are also  fully  consistent with these Ward identities.  
\end{abstract}

\setcounter{page}{0}
\setcounter{footnote}{0}
\newpage

\section{Introduction}
Many aspects of superstring theory can be captured by studying its low energy supergravity effective action. The stringy effects, however, appear in  higher-derivative and genus corrections to the supergravity. These corrections may be extracted from the most fundamental observables in the superstring theory which are the S-matrix elements \cite{Gross:1986iv,Gross:1986mw}. These objects have various hidden structure such as the Kawai-Lewellen-Tye (KLT) relations \cite{Kawai:1985xq} which connect  sphere-level S-matrix elements of closed strings to  disk-level S-matrix elements of open string states. There are similar relations between disk-level S-matrix elements of open and closed stringds and disk-level S-matrix elements of only open string states \cite{Garousi:1996ad,Hashimoto:1996bf,Stieberger:2009hq}. 
On the other hand, the S-matrix elements should expose the  dualities of superstring theory \cite{Sen:1998kr,Vafa:1997pm} through the corresponding Ward identities \cite{Garousi:2011we}. These identities can be used as generating function for the S-matrix elements, \ie they establish connections between  different elements of the scattering amplitude of  $n$ supergravitons. Calculating one  element explicitly, then all other elements of the S-matrix may be found by the Ward identities \cite{Garousi:2012gh}-\cite{Velni:2013jha}. 

Alternatively, the effective actions may be calculated directly by implementing the string dualities. The consistency of the effective action of type IIB superstring theory with S-dualiy has been used in \cite{Green:1997tv}-\cite{Basu:2007ck} to find genus and instanton correctons to the four Riemann curvature corrections to the type IIB supergravity. The consistency of non-abelian D-brane action with  T-duality has been used in \cite{Myers:1999ps} to find various commutators of the transverse scalar fields in the D-brane effective action. They have been verified by the corresponding S-matrix elements in \cite{Garousi:2000ea}. Using the consistency of the effective actions with the string dualities, some of the  eight-derivative corrections to the supergravity and   four-derivative corrections to the D-brane/O-plane world-volume effective action  have been found in  \cite{Garousi:2009dj}-\cite{Robbins:2014ara}. The complete list of the couplings  at these orders which are fully consistent with the string dualities, however,  are still lacking.

The effective actions of  type II superstring theories at the leading order of $\alpha'$ are  given by the type II supergravities which are invariant under the  string dualities. The first higher-derivative correction to these actions is at eight-derivative order. The Riemann curvature corrections to the supergravities, $t_8t_8R^4$, have been found in \cite{Gross:1986iv} from the $\alpha'$-expansion of the sphere-level S-matrix element of four graviton vertex operators. These couplings have been extended in \cite{Gross:1986mw} to include all other couplings of four NSNS states by extending the Riemann curvature to the generalized Riemann curvature, \ie
\beqa
 \bar{R}_{ab}{}^{cd}&= &R_{ab}{}^{cd}-\frac{\kappa}{\sqrt{2}}\eta_{[a}{}^{[c}\phi_{;b]}{}^{d]}+  2e^{-\phi_0/2}H_{ab}{}^{[c;d]}\labell{trans}
\eeqa
where $\bar{R}$ is the generalized Riemann curvature.
 The resulting couplings are fully consistent with the corresponding S-matrix elements. The  S-matrix elements of  four   massless   states in the type II  superstring theories have only contact terms at order $\alpha'^3$. The S-dual and T-dual Ward identities of the S-matrix elements then dictate that the NSNS couplings must be combined with the appropriate RR couplings to be consistent with these Ward identities. This guiding principle    has been used in \cite{Garousi:2013lja} to find various on-shell couplings between two RR and two gravity/B-field states  and between four RR states   at order $\alpha'^3$. 

  The dilaton term in \reef{trans} is canceled when transforming it to the string frame \cite{Garousi:2012jp,Liu:2013dna}. As a result, there is no on-shell dilaton coupling between   four NSNS fields in the string frame. 
	It has been speculated in \cite{Liu:2013dna} that there may be no dilaton couplings between all higher  NSNS fields at eight-derivative level. As we will see in this paper, however, the consistency of the  couplings $t_8t_8\bar{R}^4$  with the S-dual and T-dual Ward identities   produces various non-zero couplings between the dilaton and the RR fields in the string frame.
In this paper, we are going to examine these couplings as well as the couplings found in \cite{Garousi:2013lja}  with the explicit calculation of the sphere-level S-matrix element of two RR and two NSNS vertex operators in the RNS formalism.

The outline of the paper is as follows: We begin with the section 2 which is the detail calculations of the sphere-level S-matrix element of two RR and two NSNS vertex operators in the RNS formalism. In section 3, we compare the contact terms of these  amplitudes at order $\alpha'^3$ for the grvity/B-field with the corresponding couplings that have been found in \cite{Garousi:2013lja}. In section 4, we study the dilaton couplings. Using the T-dual and S-dual Ward identities on the S-matrix element of RR five form field strength, we find various couplings at order $\alpha'^3$ in both the string and Einstein frames.   We show that the  dilaton  couplings in the Einstein frame  are fully consistent with the corresponding S-matrix element at order $\alpha'^3$.      In section 5, we   briefly discuss our results.

\section{Scattering amplitude} 

The scattering amplitude of four RR states or 
 two RR and two NSNS states in the pure spinor formalism have been calculated in \cite{Policastro:2006vt}. In this section, we are going to calculate the scattering amplitude of two RR and two NSN states in the RNS formalism \cite{Friedan:1985ge,Kostelecky:1986xg}.
In this formalism, the tree level scattering amplitude of two RR and two NSNS states is given by the correlation function of their corresponding vertex operators on the sphere world-sheet.  Since the background superghost charge of the   sphere is $Q_{\phi}=2$, one has to choose the vertex operators in the appropriate pictures to produce the compensating   charge $Q_{\phi}=-2$.  One may choose the RR vertex operators in $(-1/2,-1/2)$  picture,  one of the NSNS vertex operators in $(-1,-1)$ and the other one in $(0,0)$ picture. The final result, should be independent of the choice of the ghost picture.

Using the above picture for the vertex operators, the scattering amplitude is given by the following correlation function \cite{Friedan:1985ge,Kostelecky:1986xg}:
\beqa
{\cal A}\sim\int \prod_{i=1}^{4} d^2 z_{i} \; \lan \prod_{j=1}^{2} V_{RR}^{(-1/2,-1/2)}(z_{j},\bar{z}_{j})V_{NSNS}^{(-1,-1)}(z_{3},\bar{z}_{3})V_{NSNS}^{(0,0)}(z_{4},\bar{z}_{4})\ran \labell{amp1}
\eeqa
where the vertex operators are\footnote{Our conventions in the string theory side set $\alpha'=2$.}
\beqa
V_{RR}^{(-1/2,-1/2)}(z_{j},\bar{z}_{j})&=&(P_{-}\Gamma_{j(n)})^{AB}:e^{-\phi(z_{j})/2}S_{A}(z_{j})e^{ik_{j}\cdot X(z_{j})}:e^{-\tphi(\bar{z}_{j})/2}\tS_{B}(\bar{z}_{j})e^{ik_{j}\cdot \tX({\bar{z}_{j}})}:\nn\\
V_{NSNS}^{(-1,-1)}(z_{3},\bar{z}_{3})&=&\veps_{3\mu\nu}:e^{-\phi(z_{3})}\psi^{\mu}(z_{3})e^{ik_{3}\cdot X(z_{3})}:e^{-\tphi(\bar{z}_{3})}\tpsi^{\nu}(\bar{z}_{3})e^{ik_{3}\cdot \tX({\bar{z}_{3}})}:\nn\\
V_{NSNS}^{(0,0)}(z_{4},\bar{z}_{4})&=&\veps_{4 \alpha \beta}:(\prt X^{\alpha}(z_{4})+ik_{4}\inn\psi\psi^{\alpha}(z_{4})) e^{ik_{4}\cdot X(z_{4})}:\nn\\&&\quad\quad\times
(\prt \tX^{\beta}(\bar{z}_{4})+ik_{4}\inn\tpsi\tpsi^{\beta}(\bar{z}_{4})) e^{ik_{4}\cdot \tX(\bar{z}_{4})}:\labell{vert}
\eeqa
where  the indices $A,B,\cdots$ are the Dirac spinor indices and  $P_-=\frac{1}{2}(1-\gamma_{11})$ is the chiral projection operator which makes the calculation of the gamma matrices to be with the full $32\times 32$ Dirac matrices of the ten dimensions.  The RR polarization tensors $\veps_1^{(n-1)},\, \veps_2^{(n-1)}$ appear in $\Gamma_{1(n)}, \, \Gamma_{2(n)}$ which are defined as 
\beq
\Gamma_{i(n)}=\frac{a_n}{n!}(F_{i})_{\mu_1\cdots\mu_n}\,\gamma^{\mu_1\cdots\mu_n}
\labell{self}
\eeq
where the $n$-form $(F_i)_{\mu_1\cdots\mu_n}= \frac{1}{2}(dC_i)_{\mu_1\cdots\mu_n} $ is  the linearized R-R field strength, and the factor $a_n=-1$     in the type IIA theory and $a_n=i$   in the type IIB theory \cite{Garousi:1996ad}. The polarization tensors of the NSNS fields are given by $\veps_3, \veps_4$. The polarization tensor is  symmetric and traceless for graviton, antisymmetric for B-field and for dilaton it is
\beqa \veps_{i}^{\mu\nu}=\frac{\phi_i}{\sqrt{8}}\left(\eta^{\mu\nu}
-k_{i}^{\mu}\ell_{i}^{\nu}-\ell_{i}^{\mu}k_{i}^{\nu}\right)\labell{dilpol}
\eeqa
where $ \ell_i $ is an auxiliary vector which satisfies $ k_i\inn\ell_i=1 $ and $ \phi_i $ is the dilaton polarization
which is one. The on-shell relations for the vertex operators are $k_i^2=0$, $k_i\inn\veps_i=0$, and $\veps_i\inn k_i=0$. The normalization of the amplitude \reef{amp1} will be fixed after fixing the conformal symmetry of the integrand.

Substituting the vertex operators \reef{vert} into \reef{amp1}, and using the fact that there is no correlation between holomorphic and anti-holomorphic for the world-sheets which have no boundary, one can separate the amplitude to the holomorphic and the anti-holomorphic  parts as   
\beqa
{\cal A} \sim (P_{-} \Gamma_{1(n)})^{AB}(P_{-} \Gamma_{2(m)})^{CD}\veps_{3 \mu \nu}\veps_{4 \alpha \beta}\int \prod_{i=1}^{4} d^2 z_{i} I_{AC}^{\mu \alpha}\otimes\tilde{I}_{BD}^{\nu \beta}\labell{amp2}
\eeqa
where the holomorphic part is   
\beqa
I_{AC}^{\mu \alpha}&=&\lan:e^{-\phi(z_{1})/2}:e^{-\phi(z_{2})/2}:e^{-\phi(z_{3})}:\ran\big[\lan:S_{A}(z_{1}):S_{C}(z_{2}):\psi^{\mu}(z_{3}):\ran\nn\\&&\times\lan:e^{ik_{1}\cdot X(z_{1})}:e^{ik_{2}\cdot X(z_{2})}:e^{ik_{3}\cdot X(z_{3})}:\prt X^{\alpha}(z_{4})e^{ik_{4}\cdot X(z_{4})}:\ran\nn\\&&+\lan:S_{A}(z_{1}):S_{C}(z_{2}):\psi^{\mu}(z_{3}):ik_{4}\inn\psi\psi^{\alpha}(z_{4}):\ran\nn\\&&\times\lan:e^{ik_{1}\cdot X(z_{1})}:e^{ik_{2}\cdot X(z_{2})}:e^{ik_{3}\cdot X(z_{3})}:e^{ik_{4}\cdot X(z_{4})}:\ran\big]\labell{right}
\eeqa
and the anti-holomorphic part $\tilde{I}_{BD}^{\nu \beta}$ is given by similar expression.

In calculating the   correlators \reef{amp2}, one needs the world-sheet propagators for the holomorphic and anti-holomorphic fields \cite{Friedan:1985ge,Kostelecky:1986xg}.  Using the  standard sphere propagators, one can easily calculate the   correlators of the bosonic fields as 
\beqa
P\equiv\lan:e^{ik_{1}\cdot X(z_{1})}:e^{ik_{2}\cdot X(z_{2})}:e^{ik_{3}\cdot X(z_{3})}:e^{ik_{4}\cdot X(z_{4})}:\ran&=&\prod_{i<j}^{4} z_{ij}^{k_{i}\cdot k_{j}}\labell{boscor}\\ 
\lan:e^{ik_{1}\cdot X(z_{1})}:e^{ik_{2}\cdot X(z_{2})}:e^{ik_{3}\cdot X(z_{3})}:\prt X^{\alpha}(z_{4})e^{ik_{4}\cdot X(z_{4})}:\ran&=&\sum_{i=1}^{3} ik_{i}^{\alpha}z_{i4}^{-1}P\nonumber\\
\lan:e^{-\phi(z_{1})/2}:e^{-\phi(z_{2})/2}:e^{-\phi(z_{3})}:\ran&=& z_{12}^{-1/4}z_{13}^{-1/2}z_{23}^{-1/2}\nn
\eeqa
where $z_{ij}=z_i-z_j$. Using the conservation of momentum and the on-shell condition $k_4\inn\veps_4=0$, one can write $\sum_{i=1}^{3} ik_{i}^{\alpha}z_{i4}^{-1}=\sum_{i=1}^{2} ik_{i}^{\alpha}z_{i4}^{-1}z_{34}^{-1}z_{3i}$. This relation will be useful later on to check that the integrand is invariant under $SL(2,{R})\times SL(2,{R})$ transformations.

To calculate the correlators involving the fermion and the spin operators, one may use 
 the   Wick-like rule for the correlation function involving an arbitrary number of fermion fields and two spin operators \cite{Liu:2001qa,Garousi:2008ge}\footnote{ See \cite{Hartl:2009yf, Hartl:2010ks,Hatefi:2014lva}, for the correlation function of fermion fields with four  spin operators.
}.
 Using this rule, one finds the following results for the fermion  correlators which appear in \reef{right}:
\beqa
&&\lan:S_{A}(z_{1}):S_{C}(z_{2}):\psi^{\mu}(z_{3}):\ran = \frac{1}{\sqrt{2}}(\gamma^{\mu}C^{-1})_{AC}z_{12}^{-3/4}z_{31}^{-1/2}z_{32}^{-1/2}\labell{ferghocor}\\
&&\lan:S_{A}(z_{1}):S_{C}(z_{2}):\psi^{\mu}(z_{3}):ik_{4}\inn\psi\psi^{\alpha}(z_{4}):\ran =  \frac{i}{2\sqrt{2}}k_{4\lambda}z_{12}^{1/4}z_{31}^{-1/2}z_{32}^{-1/2}z_{41}^{-1}z_{42}^{-1}\bigg[(\gamma^{\alpha \lambda \mu}C^{-1})_{AC}
\nn\\&&\qquad\qquad\qquad\qquad\qquad\qquad+z_{12}^{-1}z_{43}^{-1}(z_{41}z_{32}+z_{31}z_{42} )[\eta^{\mu \lambda}(\gamma^{\alpha}C^{-1})_{AC}-\eta^{\mu \alpha}(\gamma^{\lambda}C^{-1})_{AC}]\bigg]\nn
\eeqa
The fractional power of $z_{ij}$ will be converted to the integer power when the ghost correlator in  \reef{boscor} multiplied the above correlators.
 
Replacing the  correlators  \reef{ferghocor} and \reef{boscor} into the scattering amplitude \reef{amp2}, and using the on-shell conditions along with the conservation of momentum,  one can easily check that the integrand of the scattering amplitude is invariant under $SL(2,{R})\times SL(2,{R})$ transformations which is the conformal symmetry of the   $z$-plane. Fixing this symmetry by setting $z_{1}=0,z_{2}\equiv z,z_{3}=1$ and $z_{4}=\infty$, one finds the following result: 
\beqa
{\cal A}=-i\frac{\kappa^2 e^{-2\phi_0}}{8}\frac{\Gamma(-s/8)\Gamma(-t/8)\Gamma(-u/8)}{\Gamma(1+s/8)\Gamma(1+t/8)\Gamma(1+u/8)}\cK\labell{amp3}
\eeqa
where Gamma functions are the standard Gamma functions that appear in four closed string amplitude \cite{Gross:1986iv}, and the closed string kinematic factor is
\beqa
\cK&=&(P_{-} \Gamma_{1(n)})^{AB}(P_{-} \Gamma_{2(m)})^{CD}\veps_{3 \mu \nu}\veps_{4 \alpha \beta} K_{AC}^{\mu \alpha}\otimes\tilde K_{BD}^{\nu \beta}\labell{kin0}
\eeqa
   In the kinematic factor, there is   an implicit factor of delta function $\delta^{10}(k_1+k_2+k_3+k_4)$ imposing conservation of momentum. The Mandelstam variables $s=-8k_1\inn k_2$, $u=-8k_1\inn k_3$ and  $t=-8k_2\inn k_3$ satisfy $s+t+u=0$, and the kinematic factor in the holomorphic part is
\beqa
K_{AC}^{\mu \alpha} &\!\!\!\!\!=\!\!\!\!\!& \frac{1}{8}\bigg[t\bigg((k_{1}^{\alpha}+k_{2}^{\alpha})(\gamma^{\mu}C^{-1})_{AC}+k_{4}^{\mu}(\gamma^{\alpha}C^{-1})_{AC}-k_{4\lambda}\eta^{\mu \alpha}(\gamma^{\lambda}C^{-1})_{AC}\bigg)\labell{kin1}\\
&&+s\bigg(k_{1}^{\alpha}(\gamma^{\mu}C^{-1})_{AC}-\frac{1}{2}[k_{4\lambda}(\gamma^{\alpha \lambda \mu}C^{-1})_{AC}-k_{4}^{\mu}(\gamma^{\alpha}C^{-1})_{AC} +k_{4\lambda}\eta^{\mu \alpha}(\gamma^{\lambda}C^{-1})_{AC}]\bigg)\bigg]\nn
\eeqa
The kinematic factor in the anti-holomorphic part is  similar to the above expression.  We have normalized the amplitude \reef{amp3} to be consistent with the field theory couplings found in \cite{Garousi:2013lja}. The background flat metric in the Mandelstam variables and in the kinematic factor is in the string frame. That is why we have normalized the amplitude by the dilaton factor $e^{-2\phi_0}$. On the other hand,   the graviton and the dilaton have the  standard kinetic term or standard propagator only in the Einstein frame. The massless poles of the amplitude \reef{amp3} then indicates that the external gravitons in the amplitude \reef{amp3} are in the Einstein frame.

As a double check of the amplitude \reef{amp3},  one should be able to  relate this amplitude  to the product of open string amplitudes of two   spinors and two gauge bosons using the KLT prescription \cite{Kawai:1985xq}.  According to the KLT prescription, the sphere-level amplitude of four closed string states is given by
\beqa
{\cal A}&=&\frac{i}{2^9\pi} \sin(\pi k_2\inn k_3)A_{\rm open}(s/8,t/8)\otimes\tA_{\rm open}(t/8,u/8)\labell{KLT}
\eeqa
where $A_{\rm open}(s/8,t/8)$ is the disk-level scattering amplitude of four open string states in the $s-t$ channel which has been calculated in \cite{Schwarz:1982jn},
\beqa
A_{\rm open}(s/8,t/8)&=&-i\kappa e^{-\phi_0}\frac{\Gamma(-s/8)\Gamma(-t/8)}{\Gamma(1+u/8)}K \labell{open}
\eeqa
where the Mandelstam variables are the same as in the closed string amplitude. The open string kinematic factor $K$  depends on the momentum and the polarization of the external   states \cite{Schwarz:1982jn}. We have normalized the amplitudes \reef{open} and \reef{KLT} to be consistent with the normalization of the amplitude \reef{amp3}.

To find the sphere-level scattering amplitude of two RR and two NSNS states, one has to consider the open string amplitude of two spinors and two gauge bosons. The kinematic factor for this case   is \cite{Schwarz:1982jn}
 \beqa
K(u_1,u_2,\zeta_3,\zeta_4)&=&
-\frac{i}{\sqrt{2}}\bigg[\frac{1}{2} s \bar u_2\gamma \inn \zeta_3\gamma \inn(k_1+k_4)\gamma\inn\zeta_4 u_1\labell{openkin}\\
&&\qquad\quad-t\bigg(\bar u_2\gamma\inn\zeta_4 u_1 k_4\inn\zeta_3
-\bar u_2\gamma\inn\zeta_3 u_1 k_3\inn\zeta_4-\bar u_2\gamma\inn k_4 u_1 \zeta_3\inn\zeta_4\bigg)\bigg]\nn
\eeqa 
where $u_1,\, u_2$ are the spinor polarizations and $\zeta_3,\,\zeta_4$ are the gauge boson polarizations. They satisfy the following on-shell relations  
\beqa
k_{i}^{2}=0,\qquad k_{i}\inn\zeta_{i}=0,\qquad (\gamma\inn k_{i}C^{-1})_{AB}u_i^{B}=0\labell{on-shell}
\eeqa
Using these relations, and the identity
\beqa
  \bar u_2^{C}(\eta^{\lambda \mu}\gamma^{\alpha}C^{-1}-\eta^{\alpha \mu}\gamma^{\lambda}C^{-1}+\eta^{\alpha \lambda}\gamma^{\mu}C^{-1}+\gamma^{\mu}\gamma^{\lambda}\gamma^{\alpha}C^{-1})_{CA}u_1^{A}&=&(\gamma^{\alpha \lambda \mu}C^{-1})_{AC}u_1^{A}u_2^{C} \labell{trans3}\nn
\eeqa
one can write the open string kinematic factor \reef{openkin} in terms of the holomorphic   kinematic factor \reef{kin1} as 
\beqa
K(u_1,u_2,\zeta_3,\zeta_4)&=& -4i\sqrt{2}u_1^{A}u_2^{C}\zeta_{3\mu}\zeta_{4\alpha} K_{AC}^{\mu \alpha}\labell{trans2}
\eeqa
Similarly for the antiholomorphic part, \ie
\beqa
\tilde K(\tilde u_1,\tilde u_2,\tilde \zeta_3,\tilde \zeta_4)&=&-4i\sqrt{2}\tilde u_1^{B}\tilde u_2^{D}\tilde\zeta_{3\nu}\tilde\zeta_{4\beta} \tilde K_{BD}^{\nu \beta} \nn
\eeqa
Using the above relations and $\Gamma(x)\Gamma(1-x)=\pi/\sin(\pi x)$,  and  substituting the following relations in \reef{KLT}     
\beqa
\zeta_{i}^{\mu}\otimes\tilde \zeta_{i}^{\nu}&\rightarrow&\veps_{i}^{\mu\nu},\quad\quad i=3,4\nn\\
u_1^{A}\otimes \tilde u_1^{B}&\rightarrow& (P_{-} \Gamma_{1(n)})^{AB}\nn\\
u_2^{C}\otimes \tilde u_2^{D}&\rightarrow& (P_{-} \Gamma_{2(m)})^{CD}\labell{trans1}
\eeqa
one recovers the amplitude \reef{amp3}, as expected. While the open string kinematic factor \reef{openkin} is the final result for the S-matrix element of two gauge bosons and two open string spinors, the closed string kinematic  factor \reef{kin0} is not yet the final result. The external closed string states are bosons, hence, the Dirac matrices in the kinematic factor must appear in  the trace operator which should then be evaluated explicitly to find the final kinematic factor of the closed string amplitude. 

The kinematic factor \reef{kin1} has one term which contains three antisymmetric gamma matrices and all other terms contain only one gamma matrix. As a result, the closed string kinematic factor \reef{kin0} has four different terms, each one has one of the following factors:
\beqa
T_1^{\sigma \tau}&=&(P_{-} \Gamma_{1(n)})^{AB}(P_{-} \Gamma_{2(m)})^{CD}(\gamma^{\sigma}C^{-1})_{AC}(\gamma^{\tau}C^{-1})_{BD}\nn\\ 
T_2^{\sigma \beta \rho \nu}&=&(P_{-} \Gamma_{1(n)})^{AB}(P_{-} \Gamma_{2(m)})^{CD}(\gamma^{\sigma}C^{-1})_{AC}(\gamma^{\beta \rho \nu}C^{-1})_{BD}\nn\\ 
T_3^{\tau \alpha \lambda \mu }&=&(P_{-} \Gamma_{1(n)})^{AB}(P_{-} \Gamma_{2(m)})^{CD}(\gamma^{\alpha \lambda \mu}C^{-1})_{AC}(\gamma^{\tau}C^{-1})_{BD}\nn\\ 
T_4^{\alpha \lambda\mu\beta \rho \nu}&=&(P_{-}\Gamma_{1(n)})^{AB}(P_{-} \Gamma_{2(m)})^{CD}(\gamma^{\alpha \lambda \mu}C^{-1})_{AC}(\gamma^{\beta \rho \nu}C^{-1})_{BD} \labell{tr0}
\eeqa
which can be written in terms of the RR field strengths and the trace of the gamma matrices. Using the above factors,   one may then separate  the closed string kinematic factor to the following parts:
\beqa
{\cal K}={\cal K}_1+{\cal K}_2+{\cal K}_3+{\cal K}_4\labell{kin2}
\eeqa
where
\beqa
{\cal K}_1 &=& \frac{1}{256}\bigg[(t-u)^2 {k}_{4\alpha } {k}_{4\beta} {\veps}_{3\lambda \mu }{\veps}_4^{\lambda \mu} +(t-u)^2 {k}_4^{\lambda }\bigg({k}_4^{\mu } {\veps}_{3\mu \lambda } {\veps}_{4\alpha \beta} \nn\\
&&-{k}_{4\beta }({\veps}_{3\lambda \mu } {\veps}_{4\alpha \mu}+{\veps}_{3\mu \lambda } {\veps}_{4\mu \alpha})\bigg) +2 t{k}_2^{\lambda } \bigg((t-u) {k}_4^{\mu } ({\veps}_{3\mu \beta } {\veps}_{4 \alpha \lambda }+{\veps}_{3\beta \mu }{\veps}_{4\lambda \alpha}) \nn\\
&&-(t-u) {k}_{4\beta }{\veps}_{3\alpha \mu } {\veps}_{4\lambda}{}^{ \mu}+(2 t{k}_2^{\mu } {\veps}_{3\alpha\beta }-(t-u) {k}_{4\alpha } {\veps}_3{}^{\mu}{}_{\beta }){\veps}_{4\mu \lambda }\bigg)\nn\\
&&-2u {k}_1^{\lambda } \bigg((t-u) {k}_4^{\mu } ({\veps}_{3\mu \beta } {\veps}_{4\alpha\lambda }+{\veps}_{3\beta \mu }{\veps}_{4\lambda \alpha})-(t-u) {k}_{4\beta }{\veps}_{3\alpha \mu }{\veps}_{4\lambda}{}^{ \mu }\nn\\
&&-(2u {k}_1^{\mu } {\veps}_{3\alpha\beta } +(t-u) {k}_{4\alpha } {\veps}_3{}^{\mu}{}_{\beta }) {\veps}_{4\mu\lambda }
+2 t {k}_2^{\mu }{\veps}_{3\alpha \beta } ({\veps}_{4\lambda \mu}+{\veps}_{4\mu \lambda})\bigg)\bigg] T_1^{\alpha \beta}\nn\\
{\cal K}_2 &=& \frac{s}{256}\bigg[ (u-t) {k}_{4\alpha }{k}_{4\lambda } {\veps}_{3\beta \nu } {\veps }_4{}^{\beta}{}_{ \mu } +{k}_{4\lambda } \bigg((t-u){k}_4^{\beta } {\veps }_{3\beta \nu } {\veps }_{4\alpha \mu } \nn\\
&&+2(t{k}_2^{\beta }-u {k}_1^{\beta }) {\veps}_{3\alpha \nu } {\veps} _{4\beta \mu }\bigg)  \bigg]T_2^{\alpha  \lambda \mu\nu}\nn\\
{\cal K}_3 &=& \frac{s}{256}\bigg[ (u-t) {k}_{4\alpha }{k}_{4\lambda } {\veps}_{3\nu\beta } {\veps }_{4\mu}{}^{\beta } +{k}_{4\lambda } \bigg((t-u){k}_4^{\beta } {\veps }_{3\nu \beta } {\veps }_{4\mu \alpha }\nn\\
&&+2(t{k}_2^{\beta }-u {k}_1^{\beta }) {\veps}_{3\nu \alpha } {\veps}_{4\mu \beta }\bigg)  \bigg]T_3^{\alpha  \lambda \mu\nu}\nn\\
{\cal K}_4 &=& \frac{s^2}{256}\bigg[  {k}_{4\alpha } {k}_{4\beta }{\veps }_{3\rho \mu } {\veps }_{4\nu \lambda }\bigg] T_4^{\alpha \nu \rho \beta \lambda \mu }\labell{kin3}
\eeqa
Note that the tensor $T_2$ $(T_3)$ is totally antisymmetric with respect to its last three indices, hence, the indices of the NSNS momenta and  polarization tensors in $\cK_2$ $(\cK_3)$ which contract with this tensor, must be antisymmetrized. Similarly  the tensor   $T_4$ is totally antisymmetric with respect to its first and its second three indices, so the momenta and the polarization tensors in $\cK_4$ should be antisymmetrized accordingly. 

One may try to write the polarization tensors and the momenta of the NSNS states in the form  of $\veps^{[\mu}{}_{[\alpha}k^{\nu]}k_{\beta]}$ which is the generalized Riemann curvature in the momentum space. Such manipulation has been done in \cite{Policastro:2006vt} for finding the couplings of two RR and two NSNS states in the pure spinor formalism. However, we are interested in this paper in  the form of couplings which are manifestly invariant under the linear T-duality and S-duality. This form of couplings may not be  in terms of the generalized Riemann curvature.

To proceed further and write   the  kinematic factors \reef{kin3} in terms of the momenta and the polarization tensors of the external states, one has to find the explicit form of the tensors $T_1,\cdots, T_4$ in terms of the  metric $\eta_{\mu\nu}$ and the RR fields strengths $F_1,\, F_2$.
 Using the properties of the charge conjugation matrix and the Dirac matrices (see \eg appendix B. in \cite{Garousi:1996ad}), one can write the tensors  $T_1,\cdots, T_4$  as
\beqa
T_1^{\sigma \tau}&=& -\frac{(-1)^{\frac{1}{2}m(m+1)}a_{n}a_{m}}{2\,n!m!}F_{1\mu_{1}\cdots\mu_{n}}F_{2\nu_{1}\cdots\nu_{m}}\Tr(\gamma^{\sigma} \gamma^{\mu_{1}\cdots\mu_{n}}\gamma^{\tau} \gamma^{\nu_{1}\cdots\nu_{m}})\nn\\
T_2^{\sigma \beta \rho \nu} &=&-\frac{(-1)^{\frac{1}{2}m(m+1)}a_{n}a_{m}}{2\,n!m!}F_{1\mu_{1}\cdots\mu_{n}}F_{2\nu_{1}\cdots\nu_{m}}\Tr(\gamma^{\sigma} \gamma^{\mu_{1}\cdots\mu_{n}}\gamma^{\beta \rho \nu} \gamma^{\nu_{1}\cdots\nu_{m}})\nn\\
T_3^{\tau \alpha \lambda \mu } &=&-\frac{(-1)^{\frac{1}{2}n(n+1)}a_{n}a_{m}}{2\,n!m!}F_{1\mu_{1}\cdots\mu_{n}}F_{2\nu_{1}\cdots\nu_{m}}\Tr(\gamma^{\tau} \gamma^{\mu_{1}\cdots\mu_{n}}\gamma^{\alpha \lambda \mu} \gamma^{\nu_{1}\cdots\nu_{m}})\nn\\
T_4^{\alpha \lambda\mu\beta \rho \nu} &=&\frac{(-1)^{\frac{1}{2}m(m+1)}a_{n}a_{m}}{2\,n!m!}F_{1\mu_{1}\cdots\mu_{n}}F_{2\nu_{1}\cdots\nu_{m}}\Tr(\gamma^{\alpha \lambda \mu} \gamma^{\mu_{1}\cdots\mu_{n}}\gamma^{\beta \rho \nu} \gamma^{\nu_{1}\cdots\nu_{m}})\labell{tr}
\eeqa
In the chiral projection operator $P_-=\frac{1}{2}(1-\gamma_{11})$, 1 corresponds to the RR field strength $F_n$ and $\gamma_{11}$ corresponds to $F_{10-n}$ which is the magnetic dual of $F_n$ at the linear order. One may  ignore  $\gamma_{11}$ and assume that $1\leq n \leq 9$. The corresponding couplings then produce corrections to the democratic form of the supergravity \cite{Fukuma:1999jt}. 

The above traces indicate that when the difference between $n$ and $m$ is an odd number, these tensors are zero. This is what one expects because there is no couplings between the RR fields in the type IIA theory in which the RR field strengths have even rank, and the type IIB theory in which the RR field strengths have odd rank.
 When the difference between $n$ and $m$ is an even number, the traces are not zero. One can easily verify that the traces are zero for $n=m+8$. For $n=m+6$ case,  the traces in $T_1,\,T_2,\,T_3$ are zero and  $T_4$   becomes totally antisymmetric. However, the corresponding kinematic factor ${\cal K}_4$ has $k_{4\alpha}k_{4\beta}$, so the kinematic factor ${\cal K}$ is zero in this case too. Therefore, there are three cases to consider, \ie $n=m$, $n=m+2$ and $n=m+4$.

 For the case that $n=m+4$, one can easily find $T_1=0$ and $T_2=T_3$. A prescription for calculating the traces is given in the appendix A. Using it, one finds the tensors $  T_2,\, T_4$ to be
\beqa
T_2^{\sigma \beta \rho \nu} &=& -16\frac{(-1)^{n(n+1)}a_{n}^2 }{(n-4)!} F_{12}^{ \nu \rho \beta \sigma} \labell{tn=m+4}\\
T_4^{\alpha \lambda\mu\beta \rho \nu} &=& -16\frac{(-1)^{n(n+1)}a_{n}^2 }{(n-4)!}\bigg[3(\eta^{\alpha  \beta}F_{12}^{ \nu \mu \rho \lambda}+\eta^{\lambda  \beta}F_{12}^{ \nu \rho \mu \alpha}-\eta^{\mu  \beta}F_{12}^{ \nu \rho \lambda \alpha})\nn\\
&&-(n-4)(F_{12}^{ \nu \mu \rho \lambda \beta \alpha}+F_{12}^{ \nu \rho \beta \mu \alpha \lambda}-F_{12}^{ \nu \rho \beta \lambda \alpha \mu}-F_{12}^{ \rho \beta \mu \lambda \alpha \nu}+F_{12}^{\nu \beta \mu \lambda \alpha \rho}-F_{12}^{ \nu \rho \mu \lambda \alpha \beta})\bigg]\nn
\eeqa
Here   we have used   the fact that $a_{n-4}=a_{n}$, and   have used the following notation for $F_{12}$s:  
\beqa
F_{12}^{\nu \rho \beta \sigma} &\equiv& F_{1\mu_{1}\cdots\mu_{n-4}}{}^{\nu \rho \beta \sigma} F_2^{\mu_{1}\cdots\mu_{n-4}}\nn\\
F_{12}^{\nu \rho \mu \beta \lambda \alpha} &\equiv& F_{1\mu_{1}\cdots\mu_{n-5}}{}^{\nu \rho \mu \beta \lambda} F_2^{\mu_{1}\cdots\mu_{n-5}\alpha}\labell{n=m+4}
\eeqa
Note that $F_1$ is $n$-form and $F_2$ is $(n-4)$-form.  Replacing the   tensors $T_1,\cdots, T_4$ into the kinematic factor \reef{kin2}, one finds the amplitude \reef{amp3} for one RR $n$-form, one RR $(n-4)$-form  and two NSNS states. We have checked that the amplitude satisfies the Ward identity corresponding to the NSNS gauge transformations.  The kinematic factor can be further  simplified after specifying the NSNS states. As we will see in the next section, the amplitude is non-zero only when the two NSNS states are antisymmetric. For the cases that $n=m$ and $n=m+2$, the traces have been calculated in the appendix A.

To find the couplings which are produced by the amplitude \reef{amp3}, one has to expand the  gamma functions in \reef{amp3} at low energy, \ie  
\beqa
\frac{\Gamma(-s/8)\Gamma(-t/8)\Gamma(-u/8)}{\Gamma(1+s/8)\Gamma(1+t/8)\Gamma(1+u/8)}&=&-\frac{2^9}{stu}-2\zeta(3)-\frac{s^2+su+u^2}{32}\zeta(5)+\cdots\labell{expa}
\eeqa
where dots refer to higher order contact terms. The first term corresponds to the massless poles in the Feynman amplitude of two RR and two NSNS fields  which are reproduced by the supergravity couplings. We have done this calculation in the appendix B. All other terms correspond to the on-shell higher-derivative couplings of two RR and two NSNS fields in the momentum space, \ie
\beqa
\cA_c&=&-i\frac{\kappa^2e^{-2\phi_0}}{8}\left( -2\zeta(3)-\frac{s^2+su+u^2}{32}\zeta(5)+\cdots\right)\cK\labell{contact}
\eeqa
Since the above amplitude contains  only the contact terms, one has to be able to rewrite it in terms of the  RR and the NSNS field strengths.  
Moreover, the contact terms \reef{contact}  should   satisfy the T-dual and S-dual  Ward identities as well \cite{Garousi:2011we}-\cite{Velni:2012sv}. The couplings of two RR  field strengths and two Riemann curvatures/B-field strengths  at eight-derivative level   have been  found in \cite{Garousi:2013lja}  by imposing the  above  Ward identities on  the   four generalized Riemann curvature couplings \cite{Gross:1986mw}.   In the next section we will compare those  couplings  with the corresponding contact terms  in \reef{contact}.

\section{Gravity and B-field couplings} 

In this section we are going to simplify the kinematic factor in \reef{contact} for the specific NSNS states which are either graviton or B-field, and compare them with the couplings that have been found in \cite{Garousi:2013lja}. These couplings have structure $F^{(n)}F^{(n)}RR$, $F^{(n)}F^{(n)}HH$, $F^{(n)}F^{(n-2)}RH$ and $F^{(n)}F^{(n-4)}HH$ where $R$  stands for the Riemann curvature and $H$ stands for the derivative of B-field strength. The coupling with structure $F^{(n)}F^{(n-4)}RR$ has been found  to be zero. Using the explicit form of the $T_i$ tensors in \reef{tn=m+4}, we have found that $\cK=0$ when the two NSNS states are symmetric and traceless. Therefore, there is no on-shell higher-derivative coupling between one RR field strength $F^{(n)}$, one $F^{(n-4)}$ and two gravitons, as expected.

\subsection{$F^{(n)}F^{(n)}RR$} 

To find the contact terms with structure $F^{(n)}F^{(n)}RR$, one should first  simplify  the kinematic factors in \reef{kin3} when the two NSNS polarization tensors are symmetric and traceless. One should then use  the explicit form of the tensors $T_1,\cdots, T_4$  calculated in  \reef{n=m}. Using the totally antisymmetric property of the RR field strengths and taking the on-shell relations into account, one can write the kinematic factor \reef{kin2}  as   
\beqa
\cK &=& \frac{n}{2^6} \bigg[(n-1) s \bigg((n-2) s F_{12}^{\alpha \beta \lambda \mu \nu \rho} {h}_{3\lambda  \rho } {h}_{4\beta  \nu } {k}_{4\alpha } {k}_{4\mu }+F_{12}^{\alpha \beta \mu \nu} [-{h}_{3\beta  \nu } (t {k}_2^{\rho }-u {k}_1^{\rho }) \nn\\
&&  \times({h}_{4\mu  \rho } {k}_{4\alpha }-{h}_{4\alpha  \rho } {k}_{4\mu })-u {h}_{3\nu}{}^{ \rho } {k}_{4\alpha } ({h}_{4\beta  \rho } {k}_{4\mu }-{h}_{4\beta  \mu } {k}_{4\rho })+t {h}_{3\beta}{}^{  \rho } {k}_{4\mu } \nn\\
&& \times({h}_{4\alpha  \nu } {k}_{4\rho }-{h}_{4\nu  \rho } {k}_{4\alpha })]\bigg)+F_{12}^{\alpha \mu} \bigg({h}_{3\alpha  \mu } {h}_4^{\nu  \rho } [u^2 {k}_{1\nu } {k}_{1\rho }+t {k}_{2\nu } (t {k}_{2\rho }-2 u {k}_{1\rho })]\nn\\
&&+u {h}_{3\mu}{}^{\nu } (t {k}_2^{\rho }-u {k}_1^{\rho }) ({h}_{4\nu  \rho } {k}_{4\alpha }-{h}_{4\alpha  \rho } {k}_{4\nu }) +t [-{h}_{3\alpha}{}^{\nu } (t {k}_2^{\rho }-u {k}_1^{\rho })({h}_{4\nu  \rho } {k}_{4\mu }-{h}_{4\mu  \rho } {k}_{4\nu })\nn\\
&& -u {h}_3^{\nu  \rho } ({h}_{4\nu  \rho } {k}_{4\alpha } {k}_{4\mu }-{k}_{4\nu } ({h}_{4\mu  \rho } {k}_{4\alpha }+{h}_{4\alpha  \rho } {k}_{4\mu }-{h}_{4\alpha  \mu } {k}_{4\rho }))]\bigg)\bigg]\labell{FnR}
\eeqa
where $ h_3 $, $ h_4 $ are the graviton polarizations and $1\leq n\leq 9$. In   order to compare the above kinematic factor with the eight-derivative couplings found, we have to write the couplings in both cases in terms of independent variables. To this end, we have to first write the RR field strengths in the above kinematic factor in  terms of the RR potential.  Then using the conservation of momentum and the on-shell relations,  one may write it in terms of the momentum $k_1,k_2,k_3$ and  in terms of the independent Mandelstam variables  $s,u$.  Moreover, one should  write $k_3\inn h_4=-k_1\inn h_4-k_2\inn h_4$ and    $ h_4\inn k_3=- h_4\inn k_1-h_4\inn k_2$ to rewrite  the  kinematic factor in terms of   the  independent variables. The Ward identity corresponding to the gauge transformations and the symmetry of the amplitude  under the interchange of  $ 1 \leftrightarrow 2 $ and $ 3 \leftrightarrow 4 $ can easily be verified in this form. Transforming  the $F^{(n)}F^{(n)}RR$ couplings found  in \cite{Garousi:2013lja} to the momentum space and doing the same steps as above to write them in terms of the independent variables, we have found exact agreement  between \reef{FnR} and the couplings  $F^{(n)}F^{(n)}RR$ for $ n=1,2,3,4,5 $.

One can easily extend the couplings with structure  $F^{(5)}F^{(5)}RR$  to  $F^{(n)}F^{(n)}RR$ with $6\leq n\leq 9$.  Using $F^{(5)}F^{(5)}RR$ couplings, one can use the dimensional reduction on a circle, $y\sim y+2\pi$, and find the 9-dimensional couplings with structure $F^{(5)}F^{(5)}RR$   which have no Killing index. Then under the linear T-duality, these couplings transforms to the couplings with structure $F_y^{(6)}F_y^{(6)}RR$. Following \cite{Garousi:2013lja}, one can easily complete the $y$-index and find the 10-dimensional couplings  with structure $F^{(6)}F^{(6)}RR$. They produce  the correct couplings since it is impossible to have   $F^{(6)}F^{(6)}RR$ couplings in which the RR field strengths have no contraction. Repeating the above steps, one finds $F^{(n)}F^{(n)}RR$ with $6\leq n\leq 9$.

\subsection{$F^{(n)}F^{(n)}HH$} 

To find the contact terms with structure $F^{(n)}F^{(n)}HH$, one should first  simplify  the kinematic factors in \reef{kin3} when the two NSNS polarization tensors are antisymmetric. Then, one should use  the explicit form of the tensors $T_1,\cdots, T_4$  in  \reef{n=m}. Using the totally antisymmetric property of the RR field strengths and taking the on-shell relations into account, one can write the kinematic factor \reef{kin2} in this case as 
\beqa
\cK &=& \frac{1}{2^8}b_{3\alpha  \beta } \bigg[2 n (t^2+u^2) b_4^{\alpha  \beta } F_{12}^{\lambda \rho} k_{4\lambda } k_{4\rho }+4 b_4{}^{\beta}{}_{\lambda } \bigg((u-t) F_{12} (t k_2^{\lambda }-u k_1^{\lambda }) k_4^{\alpha }\nn\\
&&+n [t F_{12}^{\mu \alpha} (t k_2^{\lambda }-u k_1^{\lambda })+u^2 F_{12}^{\mu \lambda} k_4^{\alpha }] k_{4\mu }+n [t  (t F_{12}^{\lambda \rho} k_4^{\alpha }-(n-1) s F_{12}^{\lambda \mu \alpha \rho} k_{4\mu })\nn\\
&&-u (F_{12}^{\alpha \rho} (t k_2^{\lambda }-u k_1^{\lambda })+(n-1) s F_{12}^{\alpha \mu \lambda \rho} k_{4\mu })] k_{4\rho }\bigg)+n b_{4\lambda  \mu } \bigg(2  (t k_2^{\lambda }-u k_1^{\lambda })\nn\\
&& \times [-2 t F_{12}^{\mu \beta} k_4^{\alpha }+(n-1) s F_{12}^{\mu \nu \alpha \beta} k_{4\nu }]+2 u  k_4^{\alpha } [2 F_{12}^{\beta \mu} (t k_2^{\lambda }-u k_1^{\lambda })+(n-1) s\nn\\
&& \times F_{12}^{\beta \nu \lambda \mu} k_{4\nu }]+(n-1) s [ (2 t F_{12}^{\lambda \mu \beta \rho} k_4^{\alpha }+(n-2) s F_{12}^{\lambda \mu \nu \alpha \beta \rho} k_{4\nu })-2 F_{12}^{\alpha \beta \mu \rho}\nn\\
&& \times (t k_2^{\lambda}-u k_1^{\lambda })+(n-2) s F_{12}^{\alpha \beta \nu \lambda \mu \rho} k_{4\nu }] k_{4\rho }\bigg)\bigg]\labell{FnH}
\eeqa
where $ b_3 $, $ b_4 $ are the B-field polarization tensors. 
Writing the above kinematic factor and the couplings $F^{(n)}F^{(n)}HH$ found in \cite{Garousi:2013lja} in terms of the independent variables, we have found that  they are exactly identical   for $ n=1,2,3,4,5 $. Using the consistency of the couplings with the linear T-duality, one can easily extend the $F^{(5)}F^{(5)}HH$ couplings   to  $F^{(n)}F^{(n)}HH$  with $6\leq n\leq 9$.

\subsection{$F^{(n)}F^{(n-2)}HR$} 

To find the contact terms with structure $F^{(n)}F^{(n-2)}HR$, one should first  simplify  the kinematic factors in \reef{kin3} when  one of the NSNS polarization tensors is symmetric and traceless, and the other one is  antisymmetric. Then, one should use  the explicit form of the tensors $T_1,\cdots, T_4$  in  \reef{tn=m+2}. In this case, one finds 
\beqa
\cK &=& -\frac{1}{2^7}{b}_{3\alpha  \beta } \bigg[F_{12}^{\alpha  \beta} {h}_4^{\nu  \rho } [u^2 {k}_{1\nu } {k}_{1\rho }+t {k}_{2\nu } (t {k}_{2\rho }-2 u {k}_{1\rho })]+\bigg(2 u^2 F_{12}^{\nu  \rho } {h}_4{}^{\beta}{}_{\rho } {k}_4^{\alpha }\nn\\
&&+(n-2) s [(n-3) s  F_{12}^{\alpha  \beta  \nu  \rho \lambda \mu} {h}_{4\mu  \rho } {k}_{4\lambda }+2 u F_{12}^{\beta  \nu  \rho  \lambda} ({h}_{4\lambda  \rho } {k}_4^{\alpha }-{h}_4{}^{\alpha}{}_{ \rho } {k}_{4\lambda })]\bigg) {k}_{4\nu }\nn\\
&&+2 u F_{12}^{\beta  \nu } (t {k}_2^{\rho }-u {k}_1^{\rho }) ({h}_{4\nu  \rho } {k}_4^{\alpha }-{h}_4{}^{\alpha}{}_{\rho } {k}_{4\nu })-(n-2) s F_{12}^{\alpha  \beta  \nu \lambda} (t {k}_2^{\rho }-u {k}_1^{\rho }) \nn\\
&& \times({h}_{4\nu  \rho } {k}_{4\lambda }-{h}_{4\lambda  \rho } {k}_{4\nu })\bigg]\labell{Fn-2HR}
\eeqa
Writing the above kinematic factor and the couplings $F^{(n)}F^{(n-2)}HR$ found in \cite{Garousi:2013lja} in terms of the independent variables, we have found that they are exactly identical   for   $ n=3,4,5 $. 

Here also one can use the dimensional reduction on  $F^{(5)}F^{(3)}HR$ couplings and consider the 9-dimensional couplings  $F^{(5)}F^{(3)}HR$ which have no $y$-index. Then under the linear T-duality one finds  $F_y^{(6)}F_y^{(4)}HR$. Since it is impossible to have coupling $F^{(6)}F^{(4)}HR$ in which the RR field strengths have no contraction with each other, one can find all $F^{(6)}F^{(4)}HR$ couplings by completing the $y$-index in the above $F_y^{(6)}F_y^{(4)}HR$ couplings. So the couplings corresponding the  kinematic factor \reef{Fn-2HR} for $6\leq n\leq 9$ can easily be read from the $F^{(5)}F^{(3)}HR$ couplings.

\subsection{$F^{(n)}F^{(n-4)}HH$} 

To find the contact terms with structure $F^{(n)}F^{(n-4)}HH$, one should first  simplify  the kinematic factors in \reef{kin3} when    the NSNS polarization tensors are  antisymmetric. Then, one should use  the explicit form of the $T_i$  tensors    in  \reef{tn=m+4}. In this case, one finds 
\beqa
\cK &=& \frac{s}{2^8}  {b}_{3\alpha  \beta } {b}_{4\lambda  \mu }{k}_{4\rho } \bigg[2 F_{12}^{\alpha  \beta  \mu  \rho } (t {k}_2^{\lambda }-u {k}_1^{\lambda }) +2 u F_{12}^{\beta  \lambda  \mu  \rho } {k}_4^{\alpha }+(n-4) s F_{12}^{\alpha  \beta  \lambda  \mu  \rho \nu } {k}_{4\nu }\bigg]  \labell{Fn-4H}
\eeqa
Writing it in terms of the  independent variables, one finds that it is invariant under the interchange of  $ 3 \leftrightarrow 4 $, and it satisfies the Ward identity corresponding to the B-field gauge transformation. The minimum value of $n$ is 5, so we   consider $n=5$ and compare it with the couplings with structure $F^{(1)}F^{(5)}HH$.  

The $F^{(1)}F^{(5)}HH$ couplings have been found in \cite{Garousi:2013lja} by using the  dimensional reduction on the couplings with structure $F^{(2)}F^{(4)}HR$ which   have been verified by explicit calculation in the previous section.   Then under the T-duality the couplings with structure  $F^{(2)}_yF^{(4)}HR_y$ where the index $y$ is the Killing index, transform to the couplings with structure $F^{(1)}F^{(5)}_yHH_y$. One can  complete the  $y$-index to  find the   couplings with structure  $F^{(1)}F^{(5)}HH$ in the string frame\footnote{Note that in writing the field theory couplings we have used only the lowercase indices and the repeated indices are contracted with the metric $g_{\mu\nu}$.}, \ie
\beqa
S &\supset& \frac{\gamma}{\kappa^2}  \int d^{10}x\sqrt{-G}\big[ 8 F_{h,k} F_{{mnpqr},s} H_{{hpq},m} H_{{krs},n}\labell{F1F5}\\&&\qquad\qquad\qquad\quad+4 F_{h,k} F_{{kmnpq},r} H_{{mns},h} H_{{pqr},s}-2 F_{h,k} F_{{kmnpq},h} H_{{mns},r} H_{{pqr},s}
\big]\nn
\eeqa
where $\gamma=\alpha'^3\zeta(3)/2^5$. There is one extra term in the couplings that have been found in \cite{Garousi:2013lja} which is zero on-shell. Moreover, there is   an extra factor of $-1/2$ in the above action which is resulted from completing the Killing $y$-index and considering the fact that three are two B-field strengths in the couplings with structure $F^{(1)}F^{(5)}_yHH_y$.
Transforming the above action to the momentum space, and writing the couplings in terms of the independent variables, we have found exact agreement with the kinematic factor in \reef{Fn-4H}.

One can use the dimensional reduction on  above  $F^{(1)}F^{(5)}HH$ couplings and consider the 9-dimensional couplings  $F^{(1)}F^{(5)}HH$ which have no $y$-index. Then under the linear T-duality one finds  $F_y^{(2)}F_y^{(6)}HH$. Completing the $y$-index, one finds all couplings in which the RR field strengths have contraction with each other. In this case, however,   it is possible to have coupling $F^{(2)}F^{(6)}HH$ in which the RR field strengths have no contraction with each other. We add all  such couplings with unknown coefficients, and constrain them to be consistent with the kinematic factor \reef{Fn-4H}.  We find the following result for $F^{(2)}F^{(6)}HH$ couplings: 
\beqa
S &\supset& \frac{\gamma}{\kappa^2}  \int d^{10}x\sqrt{-G}\big[8 F_{ht,k} F_{mnpqrt,s} H_{hpq,m} H_{krs,n}+8 F_{hm,n} F_{nkpqrs,t} H_{hkp,q} H_{mrs,t}\nn\\&&\qquad\qquad\qquad+4 F_{ht,k} F_{kmnpqt,r} H_{mns,h} H_{pqr,s}-2 F_{ht,k} F_{kmnpqt,h} H_{mns,r} H_{pqr,s}\big]\labell{new1}
\eeqa
The first term and the couplings in the second line are the couplings that can be read from the T-duality of the couplings \reef{F1F5}. The second term is the coupling in which the RR field strengths have no contraction with each other, hence, it could not be read from the T-duality of \reef{F1F5}.

The couplings with structure $F^{(n-4)}F^{(n)}HH$ for $n>6$ can easily be read from the T-duality of the couplings \reef{new1} because it is impossible to have such couplings in which the RR field strengths have no contraction with each other. The result is 
\beqa
S &\!\!\!\supset\!\!\!& \frac{1}{(n-5)!}\frac{\gamma}{\kappa^2}  \int d^{10}x\sqrt{-G}\big[8 F_{ha_1\cdots a_{n-5},k} F_{mnpqra_1\cdots a_{n-5},s} H_{hpq,m} H_{krs,n}\nn\\&&-2 F_{ha_1\cdots a_{n-5},k} F_{kmnpqa_1\cdots a_{n-5},h} H_{mns,r} H_{pqr,s}+4 F_{ha_1\cdots a_{n-5},k} F_{kmnpqa_1\cdots a_{n-5},r} H_{mns,h} H_{pqr,s}\nn\\&&\qquad\qquad\qquad\qquad\qquad+8(n-5) F_{hma_1\cdots a_{n-6},n} F_{nkpqrsa_1\cdots a_{n-6},t} H_{hkp,q} H_{mrs,t}\big]\labell{new2}
\eeqa
The number of indices $a_1\cdots a_{n-m}$ in the RR field strengths is such that the total number of the indices of $F^{(n)}$ must be $n$. For example, $F_{nkpqrsa_1\cdots a_{n-6},t}$ is  $F_{nkpqrs,t}$ for $n=6$ and is zero for $n<6$. We have checked that the above couplings are consistent with the kinematic factor \reef{Fn-4H} for $n\geq 5$.

\section{Dilaton couplings} 

In this section we are going to simplify the kinematic factor in \reef{contact} for the cases that one or both of the  NSNS states are dilatons.  One has to use \reef{dilpol} for the dilaton polarization. There are three cases to consider. The first case is when one of the polarizations is the dilaton and the other one is antisymmetric. The kinematic factor in this case is non-zero for $n=m+2$. So the non-zero couplings should have structure $F^{(n)}F^{(n-2)}H\phi$ where $\phi$ stands for the second derivatives of the dilaton. The second case is when one of the polarization tensors is the dilaton and the other one is symmetric and traceless. The kinematic factor in this case is non-zero for $n=m$. So the non-zero couplings in this case should have structure $F^{(n)}F^{(n)}R\phi$. The third case is when both of the NSNS polarizations are the dilatons. The kinematic factor in this case is also non-zero for $n=m$. The non-zero couplings should have the structure $F^{(n)}F^{(n)}\phi\phi$. In all cases, we have found that the auxiliary vector $\ell$ of the dilaton polarization \reef{dilpol} is canceled in the kinematic factors, as expected. Let us begin with the first case.

\subsection{$F^{(n)}F^{(n-2)}H\phi$}

Replacing the tensors $T_i$ for $ n=m+2 $ case which are calculated in \reef{tn=m+2}, into \reef{kin3}, one finds the following result for the kinematic factor \reef{kin2} when one of the polarization tensors is antisymmetric and the other is the dilaton polarization \reef{dilpol}: 
\beqa
\cK &=& -\frac{(n-7) s-2  t}{2^8 \sqrt{2} }\phi_3 {b}_{4\alpha  \beta } {k}_{4\nu } \bigg[2  F_{12}^{\beta \nu } (t {k}_2^{\alpha }-u  {k}_1^{\alpha })+(n-2) s  F_{12}^{\alpha \beta \nu \lambda} {k}_{4\lambda }\bigg] \labell{FnHD}
\eeqa
 As has been discussed already, the external states in the contact terms \reef{contact} are in the Einstein frame whereas the background metric $\eta_{\mu\nu}$ is in the string frame. Hence, to find the appropriate couplings in a specific frame, one has  to either transform the external graviton states to the string frame   or transform the background metric $\eta_{\mu\nu}$ to the Einstein frame, \ie  $e^{\phi_0/2}\eta_{\mu\nu}$.  We choose the latter transformation to rewrite the contact terms \reef{contact} in the Einstein frame.  

 Since the couplings are in the Einstein frame, the natural question is whether the string frame couplings $ F^{(n)} F^{(n-2)} H R $, produce all the couplings  $ F^{(n)} F^{(n-2)} H \phi $ in the Einstein frame? In fact the transformation of the Riemann curvature from the string frame to the Einstein frame is \cite{Garousi:2012jp}
\beqa
 {R}_{ab}{}^{cd} & \Longrightarrow & R_{ab}{}^{cd}-\frac{\kappa}{\sqrt{2}}\eta_{[a}{}^{[c}\phi_{;b]}{}^{d]}+\cdots\labell{transf}
\eeqa
where $\phi$ is the perturbation of dilaton, \ie $\Phi=\phi_0+\sqrt{2}\kappa \phi$. In above equation dots refer to the terms with two dilatons in which we are not interested.  We have transformed the string frame couplings $ F^{(n)} F^{(n-2)} H R $ to the Einstein frame and found that the resulting  $ F^{(n)} F^{(n-2)} H \phi $ couplings are not consistent with the kinematic factor in \reef{FnHD}.   This indicates that there must be some new couplings with structure  $ F^{(n)} F^{(n-2)} H \phi $ in the string frame. The combination of these couplings and the couplings with structure  $ F^{(n)} F^{(n-2)} H R $, then must be consistent with \reef{FnHD} when transforming them to the Einstein frame. This constraint can be used to find the dilaton couplings. We will find the couplings in both string and Einstein frames for $n\leq 5$. In section 5, we extend the string frame couplings to $1\leq n \leq 9$.

The new couplings in the string frame must  be consistent with the linear T-duality. So we will first find the new couplings for the case of $n=5$ by using the consistency with \reef{FnHD} and then find the couplings for other values of $n\leq4$ by using the consistency with the linear T-duality. To find the string frame couplings with structure  $ F^{(5)} F^{(3)} H \phi $,  we  consider all possible on-shell contractions of terms with structure $ F^{(5)} F^{(3)} H \phi $ with unknown coefficients. This can be performed using the new field theory motivated package for the Mathematica "xTras" \cite{CS}. Transforming the combination of these couplings and the couplings with structure  $ F^{(5)} F^{(3)} H \phi $  found in \cite{Garousi:2013lja}, to the Einstein frame and constraining them to be consistent with the kinematic factor \reef{FnHD}, one finds some relations between the unknown coefficients. Replacing these relations into the general couplings with structure $ F^{(5)} F^{(3)} H \phi $, one finds the following couplings in the string frame:   
\beqa
S &\supset& - \frac{4 }{3} \frac{\gamma}{\kappa^2}\int d^{10}x  \sqrt{-G} \big[ 6 H_{h}{}_{rs}{}_{,m} F_{knqrs}{}_{,p} 
F_{mnp}{}_{,q} + 3 H_{h}{}_{rs}{}_{,m} F_{knprs}{}_{,q} 
F_{mnp}{}_{,q} \labell{mF5F3} \\ 
&&\qquad\qquad\qquad\qquad-  H_{h}{}_{rs}{}_{,k} F_{mnprs}{}_{,q} 
F_{mnp}{}_{,q}  - 6 F_{knprs}{}_{,q} F_{mnp}{}_{,q} H_{hm}{}_{r}{}_{,s} \nn \\ 
&&\qquad\qquad\qquad\qquad+ 6 
F_{mnp}{}_{,q} F_{knpqr}{}_{,s} H_{hm}{}_{r}{}_{,s} + 3 
F_{h}{}_{mn}{}_{,p} F_{mnpqr}{}_{,s} H_{k}{}_{qr}{}_{,s}
\big]\Phi_{,hk}\nn
\eeqa
 Plus some other terms which contains some of the unknown coefficients. However, these terms vanish when we write the field strengths in terms of the corresponding potentials and use the on-shell relations to write the result in terms of the independent variables. That means these terms are canceled using the Bianchi identities for the RR and for the B-field strengths and the on-shell relations. As a result, these terms can safely be set to zero. We refer the reader to section 4.3 in which similar calculation has been done in more details. Since we have considered all contractions without fixing completely the Bianchi identities, the above couplings are  unique up to using the Bianchi identities. That is, one may find another action which is related to the above action by using the Bianchi identities and the on-shell relations.

Having found the string frame couplings  $ F^{(n)} F^{(n-2)} H \phi $ for $n=5$ in \reef{mF5F3}, we now apply the T-duality transformations on them to find the corresponding couplings for other $n$. We use the dimensional reduction on the couplings \reef{mF5F3}  and find the couplings with structure $F^{(5)}_yF^{(3)}_y H\phi$. Under the linear T-duality transformations, the RR field strength $F^{(n)}_y$ transforms to $F^{(n-1)}$ with no Killing index, the B-field with no $y$-index is invariant and the perturbation of dilaton transforms as (see \eg  \cite{Garousi:2013lja})
\beqa
\phi\rightarrow \phi-\frac{1}{\sqrt{2}}h_{yy}\labell{delT}
\eeqa
where $h_{\mu\nu}$ is the metric perturbation, \ie $g_{\mu\nu}=\eta_{\mu\nu}+2\kappa h_{\mu\nu}$. The couplings in $F^{(5)}_yF^{(3)}_y H\phi$ corresponding to the second term above should be canceled with  the couplings  with structure $F^{(4)}F^{(2)} HR_{yy}$.  The terms corresponding to the first term in \reef{delT} have structure $F^{(4)}F^{(2)}H\phi$.   The result for all terms in \reef{mF5F3} is the following couplings in the string frame:
\beqa
S &\supset& - 4  \frac{\gamma}{\kappa^2}\int d^{10}x  \sqrt{-G} \big[ 2 H_{h}{}_{rs}{}_{,m} F_{kqrs}{}_{,p} 
F_{mp}{}_{,q} + 2 H_{h}{}_{rs}{}_{,m} F_{kprs}{}_{,q} 
F_{mp}{}_{,q} \labell{mF4F2}\\ 
&&\qquad\qquad\qquad\qquad-  H_{h}{}_{rs}{}_{,k} F_{mprs}{}_{,q} 
F_{mp}{}_{,q}  - 4 F_{kprs}{}_{,q} F_{mp}{}_{,q} H_{hm}{}_{r}{}_{,s} \nn \\ 
&&\qquad\qquad\qquad\qquad+ 4 
F_{mp}{}_{,q} F_{kpqr}{}_{,s} H_{hm}{}_{r}{}_{,s} - 2
F_{h}{}_{m}{}_{,p} F_{mpqr}{}_{,s} H_{k}{}_{qr}{}_{,s}
\big]\Phi_{,hk}\nn
\eeqa
 The transformation of the combination of the above couplings and the couplings with structure $F^{(4)}F^{(2)} HR$ which have been found in \cite{Garousi:2013lja}, to the Einstein frame is the following:
\beqa
S &\supset& \frac{\gamma}{\kappa^2}\int d^{10}x \sqrt{-G} e^{-\phi_0}\big[-2  F_{hm,n} F_{mnpq,r} H_{kpq,r} \Phi_{,hk}-2  F_{hm,n} F_{nkpq,r} H_{mpq,r} \Phi_{,hk}\nn\\&&+6  F_{hm,n} F_{npqr,k} H_{mpq,r} \Phi_{,hk}-8  F_{hm,n} F_{mkpq,r} H_{npq,r} \Phi_{,hk}+12  F_{hm,n} F_{mkpr,q} H_{npq,r} \Phi_{,hk}\nn\\&&+\frac{2}{3}  F_{hm,n} F_{mkpq,r} H_{kpq,r} \Phi_{,hn}+12  F_{hm,n} F_{mnqr,p} H_{hkq,r} \Phi_{,kp}-6  F_{hm,n} F_{mnqr,p} H_{hqr,k} \Phi_{,kp}\nn\\&&+6  F_{hm,n} F_{mkqr,p} H_{hqr,n} \Phi_{,kp}\big]\labell{EF4F2HD}
\eeqa
We have checked that it  is consistent with \reef{FnHD} for $n=4$. In writing the above result we have used the Bianchi identities and   the on-shell relations  to simplify the couplings. However, one may still use these  identities to rewrite the above couplings in a simpler form. It would be interesting to find the minimum number of terms in which  the Bianchi identities have been used completely. 

Having found the string frame couplings  $ F^{(n)} F^{(n-2)} H \phi $ for $n=4$ in \reef{mF4F2}, we now apply the  T-duality transformations on them to find the corresponding couplings for  $n=3$. To find the couplings with structure $F^{(3)}F^{(1)}H\phi$, one has to use the dimensional reduction on the couplings \reef{mF4F2} and find the couplings with structure $F^{(4)}_yF^{(2)}_y H\phi$. Then under the linear  T-duality transformations, they transform to the couplings with structure $F^{(3)}F^{(1)} H\phi$. In the string frame they are given by 
\beqa
S &\supset& - 8  \frac{\gamma}{\kappa^2}\int d^{10}x \sqrt{-G} \Phi_{,hk}\big[  H_{h}{}_{rs}{}_{,m} F_{krs}{}_{,q} 
F_{m}{}_{,q} -  H_{h}{}_{rs}{}_{,k} F_{mrs}{}_{,q} 
F_{m}{}_{,q}\labell{mF3F1} \\ 
&&\qquad\qquad\qquad\qquad\qquad  - 2 F_{krs}{}_{,q} F_{m}{}_{,q} H_{hm}{}_{r}{}_{,s} + 2 
F_{m}{}_{,q} F_{kqr}{}_{,s} H_{hm}{}_{r}{}_{,s} + 
F_{h}{}_{}{}_{,p} F_{pqr}{}_{,s} H_{k}{}_{qr}{}_{,s}
\big]\nn
\eeqa
 We have checked that the transformation of the combination of the above couplings and the couplings with structure $F^{(3)}F^{(1)} HR$, to the Einstein frame produces exactly the couplings which are consistent with \reef{FnHD} for $n=3$.

\subsubsection{Consistency with the S-duality}

The dilaton couplings in \reef{mF4F2} are in the type IIA theory whereas the couplings in \reef{mF5F3} and \reef{mF3F1} are in the type IIB theory. The effective action in the type IIB theory should be invariant under the S-duality, as a result, the couplings  in \reef{mF5F3} and \reef{mF3F1} should be consistent with the S-duality. The standard S-duality transformations are in the Einstein frame, so we have to transform the couplings with structure $F^{(n)}F^{(n-2)} HR$ and  $F^{(n)}F^{(n-2)} H\phi$ to the Einstein frame and then study their compatibility with the S-duality for $n=5,3$.

The extension of the couplings in \reef{F1F5} to the S-duality invariant form has been found in \cite{Garousi:2013lja}. Apart from the overall dilaton factor in the Einstein frame which is extended to the $SL(2,{Z})$ invariant Eisenstein series $E_{3/2}$, each coupling should be extended to the   $SL(2,{R})$ invariant form, \eg
the first term in \reef{F1F5} is extended to 
\beqa
\cH^T_{h q r,m}\cM_{,nk}\cN^{-1}\cM_0\cH_{m p s,k}&=&2F_{n,k}(H_{h q r,m} H_{m p s,k}-e^{2\phi_0}F_{h q r,m} F_{m p s,k})\labell{ex}\\
&&+\sqrt{2}\kappa\phi_{,nk}(H_{h q r,m} F_{m p s,k}+F_{h q r,m} H_{m p s,k})+\cdots\nonumber
\eeqa
where dots refer to the terms with non-zero axion background in which we are not interested. We refer the interested reader to \cite{Garousi:2013lja} for the definitions of $\cH,\,\cM$ and $\cN$. The terms in \reef{F1F5} correspond to the first term in the above $SL(2,{R})$ invariant set. 
The terms in the S-duality invariant action which correspond to the second line above have structure $F^{(5)}F^{(3)}H\phi$. We have checked explicitly that these terms are reproduced exactly by transforming the couplings \reef{mF5F3} and the couplings with structure  $F^{(5)}F^{(3)} HR$ (see eq.(35) in \cite{Garousi:2013lja}) to the Einstein frame. In other words, these couplings are fully consistent with the kinematic factor \reef{FnHD} for $n=5$. The couplings corresponding to the last term in \reef{ex} are the S-duality prediction for  four RR couplings with structure $F^{(5)}F^{(1)}F^{(3)}F^{(3)}$.

The transformation of the couplings \reef{mF3F1} and the couplings  with structure $F^{(3)}F^{(1)} HR$ (see eq.(37) in \cite{Garousi:2013lja}) to the Einstein frame, produces the following couplings with structure $F^{(3)}F^{(1)} H\phi$ in the Einstein frame: 
\beqa
S &\supset& \frac{\gamma}{\kappa^2}\int d^{10}x \sqrt{-G} e^{-\phi_0/2}\big[-\frac{4}{3}  F_{h,m} F_{nkp,q} H_{nkp,q} \Phi_{,hm}\labell{F3F1}\\&&+8  F_{h,m} F_{kpq,n} H_{mkp,q} \Phi_{,hn}+4  F_{h,m} F_{nkp,q} H_{mkp,q} \Phi_{,hn}+4  F_{h,m} F_{mkp,q} H_{nkp,q} \Phi_{,hn}\nn\\&&-8  F_{h,m} F_{npq,k} H_{hpq,m} \Phi_{,nk}-16  F_{h,m} F_{hpq,n} H_{mkp,q} \Phi_{,nk}+8  F_{h,m} F_{hpq,n} H_{mpq,k} \Phi_{,nk}\big]\nn 
\eeqa
which are consistent with the kinematic factor \reef{FnHD} for $n=3$. Using the on-shell relations, one finds that the above amplitude is invariant under the transformation 
\beqa
F^{(3)}\longrightarrow H&;& H\longrightarrow -F^{(3)}
\eeqa
It is also invariant under the following transformation:
\beqa
F^{(1)}\longrightarrow d\Phi&;&d\Phi\longrightarrow -F^{(1)}
\eeqa
 Using these properties, one should extended the amplitude \reef{F3F1} to the S-duality invariant form.

The dilaton factor in \reef{F3F1} can be rewritten as $e^{-3\phi_0/2}\times e^{\phi_0}$. The first factor is extended to the $SL(2,{Z})$ invariant function $E_{3/2}$ after including the one-loop result and the nonperturbative   effects \cite{Green:1997tv}. The second factor combines with the dilaton  and the RR scalar to produce the  following $SL(2,{R})$ invariant term:
\beqa
e^{\phi_0}(F_{h,k}\Phi_{,mn}-F_{m,n}\Phi_{,hk})
\eeqa
Using the standard $SL(2,{R})$ transformation of the dilaton and the RR scalar, \ie $\tau\longrightarrow \frac{p\tau+q}{r\tau+s}$ where $\tau=C+ie^{-\Phi}$, one finds the above term is invariant under the $SL(2,{R})$ transformation\footnote{It has been observed in \cite{Kamani:2013tv} that $e^{\Phi}F^{(1)}\wedge d\Phi$ is invariant under the ${Z}_2$ subgroup of the $SL(2,{R})$ group.}. The RR two-form and the B-field should appear in the following $SL(2,{R})$ invariant term:
\beqa
\cH^T_{mnq}{}_{,p}\cN \cH_{mnp}{}_{,q}&=&H_{mnq}{}_{,p} F_{mnp}{}_{,q}- F_{mnq}{}_{,p} H_{mnp}{}_{,q}
\eeqa
Therefore, the $SL(2,{Z})$ invariant extension of the action \reef{F3F1} has no coupling other than $F^{(3)}F^{(1)} H\phi$.

\subsection{$F^{(n)}F^{(n)}R\phi$}

Replacing the tensors $T_i$ for $ n=m $ case which are calculated in \reef{tn=m}, into \reef{kin3}, one finds the following result for the kinematic factor \reef{kin2} when one of the polarization tensors is symmetric and traceless  and the other one is the dilaton polarization \reef{dilpol}: 
\beqa
\cK &=& \frac{\phi_3}{2^7 \sqrt{2}  }(n-5)  \bigg[F_{12} {h}_4^{\mu  \nu } \bigg(u^2 {k}_{1\mu } {k}_{1\nu }+t {k}_{2\mu } (t {k}_{2\nu }-2 u {k}_{1\nu })\bigg)\nn\\
&& +n s \bigg((n-1) s F_{12}^{\alpha \beta \mu \nu} {h}_{4\beta  \nu } {k}_{4\alpha } {k}_{4\mu }-F_{12}^{\alpha \mu} (t {k}_2^{\nu }-u {k}_1^{\nu }) ({h}_{4\mu  \nu } {k}_{4\alpha }-{h}_{4\alpha  \nu } {k}_{4\mu })\bigg)\bigg]\labell{FnRD}
\eeqa
 The kinematic factor is zero for $n=5$. So there is no higher-derivative coupling between two $F^{(5)}$, one graviton and one dilaton in the Einstein frame. This result is  consistent with  the S-duality because $F^{(5)}$ and the graviton in the Einstein frame are invariant under the S-duality whereas the dilaton is not invariant under the S-duality.

Now we are going to find the couplings with structure $ F^{(n)} F^{(n)} R \phi $ for $ n=1,2,3,4,5 $ in the string frame.  To this end,  we have to first transform the string frame couplings with structure $ F^{(n)} F^{(n)} R R $ which have been found in \cite{Garousi:2013lja}, to the Einstein frame. If they do not produce the kinematic factor \reef{FnRD}, then one has to consider new couplings with structure $ F^{(n)} F^{(n)} R \phi $. So let us begin with the case of $n=5$.  Transforming  the couplings with structure $ F^{(5)} F^{(5)} R R $ (see eq.(27) in \cite{Garousi:2013lja}) to the Einstein frame, we have found that they produce the couplings with structure $ F^{(5)} F^{(5)} R \phi $ which are not zero, \ie they are not consistent with \reef{FnRD}. As a result, one has to consider new couplings in the string frame with structure $ F^{(5)} F^{(5)} R \phi $ to cancel them. We consider all such on-shell couplings with unknown coefficients, and constrain them to cancel the above  $ F^{(5)} F^{(5)} R \phi $ couplings.  This constraint produces some relations between the coefficients. Replacing them into the general couplings with structure $ F^{(5)} F^{(5)} R \phi $, one finds the following couplings: 
\beqa
S &\supset& - \frac{4 }{3} \frac{\gamma}{\kappa^2}\int d^{10}x \sqrt{-G} \big[ - R_{h}{}_{mnp} \
F_{k}{}_{qrst}{}_{,n} F_{mpqrs}{}_{,t} -  R_{h}{}_{mnp} F_{mqrst}{}_{,p} F_{kn}{}_{qrs}{}_{,t} \labell{mF5RD}\\ 
&& + 3 R_{h}{}_{mnp} F_{mpqrt}{}_{,s} \
F_{kn}{}_{qrs}{}_{,t} - 3 R_{h}{}_{mnp} \
F_{mpqrs}{}_{,t} F_{kn}{}_{qrs}{}_{,t} + R_{h}{}_{m}{}_{k}{}_{n} F_{npqrt}{}_{,s} \
F_{m}{}_{pqrs}{}_{,t} \big]\Phi_{,hk}\nn
\eeqa
Plus some other terms which contains some of the unknown coefficients. However, these terms vanish when we write the field strengths in terms of the corresponding potentials and use the on-shell relations to write the result in terms of the independent variables.   As a result, these terms can safely be set to zero. We refer the reader to section 4.3 in which similar calculation has been done in more details.

Using the above   couplings in the string frame, one can perform   the  dimensional reduction on a circle and finds  the couplings with structure $F^{(5)}_yF^{(5)}_y R\phi$. Under the linear T-duality, they produce the following couplings with structure $F^{(4)}F^{(4)} R\phi$ in the type IIA theory:
\beqa
S &\supset& -  4  \frac{\gamma}{\kappa^2}\int d^{10}x \sqrt{-G} \big[  R_{h}{}_{mnp} \
F_{k}{}_{rst}{}_{,n} F_{mprs}{}_{,t} +  R_{h}{}_{mnp} F_{mrst}{}_{,p} F_{kn}{}_{rs}{}_{,t} \labell{mF4RD}\\ 
&& + 2 R_{h}{}_{mnp} F_{mprt}{}_{,s} \
F_{kn}{}_{rs}{}_{,t} - 3 R_{h}{}_{mnp} \
F_{mprs}{}_{,t} F_{kn}{}_{rs}{}_{,t} + R_{h}{}_{m}{}_{k}{}_{n} F_{nprt}{}_{,s} \
F_{m}{}_{prs}{}_{,t} \big]\Phi_{,hk}\nn
\eeqa
The transformation of the above couplings and the couplings with structure $F^{(4)}F^{(4)} RR$  (see eq.(26) in \cite{Garousi:2013lja}), to the Einstein frame produces the following couplings:
\beqa
S &\!\!\!\supset\!\!\!& \frac{\gamma}{\kappa^2}\int d^{10}x \sqrt{-G} e^{-3\phi_0/2}\big[3  F_{mkqr,s} F_{npqr,s} R_{hnkp} \Phi_{,hm}- F_{mqrs,k} F_{npqr,s} R_{hnkp} \Phi_{,hm}\labell{EF4F4RD}\\&&-2  F_{mkqr,s} F_{npqs,r} R_{hnkp} \Phi_{,hm}- F_{mkqr,s} F_{nqrs,p} R_{hnkp} \Phi_{,hm}- F_{kpqs,r} F_{npqr,s} R_{hnmk} \Phi_{,hm}\big]\nn
\eeqa
which are exactly consistent with \reef{FnRD} for $n=4$. 

The  couplings with structure $F^{(3)}F^{(3)} R\phi$ in the string frame can be found from the   couplings \reef{mF4RD} by applying the dimensional reduction and finding the terms with structure $F^{(3)}_yF^{(3)}_yR\phi$. Then under T-duality they produce the following couplings   in the type IIB theory:
\beqa
S &\supset& -  8  \frac{\gamma}{\kappa^2}\int d^{10}x \sqrt{-G} \big[ - R_{h}{}_{mnp} \
F_{k}{}_{st}{}_{,n} F_{mps}{}_{,t} -  R_{h}{}_{mnp} F_{mst}{}_{,p} F_{kn}{}_{s}{}_{,t} \labell{mF3RD}\\ 
&& +  R_{h}{}_{mnp} F_{mpt}{}_{,s} \
F_{kn}{}_{s}{}_{,t} - 3 R_{h}{}_{mnp} \
F_{mps}{}_{,t} F_{kn}{}_{s}{}_{,t} + R_{h}{}_{m}{}_{k}{}_{n} F_{npt}{}_{,s} \
F_{m}{}_{ps}{}_{,t} \big]\Phi_{,hk}\nn
\eeqa
We have checked that the transformation of the above couplings and the couplings with structure $F^{(3)}F^{(3)} RR$ (see eq.(20) in \cite{Garousi:2013lja}), to the Einstein frame are exactly consistent with \reef{FnRD} for $n=3$. The S-duality transformations of these couplings   are    discussed in the next section.

Having found the couplings with structure $F^{(3)}F^{(3)}R\phi$ in \reef{mF3RD}, we now construct the couplings  with structure $F^{(2)}F^{(2)}R\phi$ in the type IIA theory.  Under the dimensional reduction on the above couplings,  the couplings  with structure  $F^{(3)}_yF^{(3)}_yR\phi$  produce the following couplings under  the T-duality:
\beqa
S &\supset& -  8  \frac{\gamma}{\kappa^2}\int d^{10}x \sqrt{-G} \big[  R_{h}{}_{mnp} \
F_{k}{}_{t}{}_{,n} F_{mp}{}_{,t} +  R_{h}{}_{mnp} F_{mt}{}_{,p} F_{kn}{}_{}{}_{,t} \labell{mF2RD}\\ 
&&\qquad\qquad\qquad\qquad- 3 R_{h}{}_{mnp} \
F_{mp}{}_{,t} F_{kn}{}_{}{}_{,t} + R_{h}{}_{m}{}_{k}{}_{n} F_{nt}{}_{,s} \
F_{m}{}_{s}{}_{,t} \big]\Phi_{,hk}\nn
\eeqa
Note that  in the first term in the second line of \reef{mF3RD} there is no contraction between the two RR field strengths. Hence, this term does not produce coupling with structure $F^{(3)}_yF^{(3)}_yR\phi$. That is why this term does not appear in \reef{mF2RD}. The transformation of the above couplings and the couplings with structure $F^{(2)}F^{(2)} RR$  (see eq.(21) in \cite{Garousi:2013lja}), to the Einstein frame is given by the following couplings:
\beqa
S &\supset& 12\frac{\gamma}{\kappa^2}\int d^{10}x \sqrt{-G} e^{-\phi_0/2}\big[  F_{hm,n} F_{kp,q} R_{mkpq} \Phi_{,hn}-  F_{hm,n} F_{kp,q} R_{mnpq} \Phi_{,hk}\labell{EF2F2RD}\\&&\qquad\qquad\qquad\qquad\qquad-  F_{hm,n} F_{kp,n} R_{mkpq} \Phi_{,hq}-  F_{hm,n} F_{nk,p} R_{mkpq} \Phi_{,hq}\big]\nn
\eeqa
which are exactly consistent with \reef{FnRD} for $n=2$. 

There is no contraction between the RR field strengths in \reef{mF2RD}, hence, the dimensional  reduction on a circle does not produce couplings with structure  $F^{(2)}_yF^{(2)}_yR\phi$. As a result, the linear T-duality indicates that there is no new coupling with structure  $F^{(1)}F^{(1)}R\phi$ in the string frame.  We have checked that the transformation  of the couplings with structure $F^{(1)}F^{(1)} RR$ (see eq.(22) in \cite{Garousi:2013lja}), to the Einstein frame are   consistent with \reef{FnRD} for $n=1$. In fact both are zero in this case.

\subsubsection{Consistency with the S-duality}

The Einstein frame couplings  $F^{(n)}F^{(n)}R\phi$ for $n=1,3,5$ are in the type IIB theory, so   they should be consistent with the S-duality. We have seen that for $n=1,5$ the couplings are zero which are consistent with the S-duality because it is impossible to construct the $SL(2,{R})$ invariant term from one dilaton or from one dilaton and two RR scalars. Note that the dilaton and the axion in   $E_{3/2}$ are constant, so we can not consider the derivative of  $E_{3/2}$ which produces $\partial\phi e^{-3\phi/2}$ at weak coupling. In fact the contact terms in \reef{contact} represent the couplings of four quantum states in the presence of constant dilaton background. It is totally nontrivial  to extend the amplitude \reef{amp3} to non-constant dilaton background, \ie it is not trivial to  take into account the derivatives of the dilaton background. That amplitude would produce higher-point functions.

The  couplings with structure $F^{(3)}F^{(3)}R\phi$, however, are not zero. That means it is possible to construct the  $SL(2,{R})$ invariant couplings  which contains one dilaton and two RR 2-forms. In fact the S-duality invariant couplings which include such couplings have been constructed in \cite{Garousi:2013lja}, \ie
\beqa
S &\supset&\frac{\gamma}{\kappa^2 } \int d^{10}x E_{3/2}\sqrt{-G}\big[ 4 \cH^T_{h q r,n} \cM_{,rm}\cH_{k n p,h} R_{m p k q}-4   \cH^T_{n p r,h}\cM_{,mk} \cH_{h n q,r} R_{k q m p}\nonumber\\&&-4   \cH^T_{n p q,m} \cM_{,qh}\cH_{m p r,k} R_{k r h n}+4   \cH^T_{n p q,h} \cM_{,qm}\cH_{k n r,h} R_{m r k p}-2   \cH^T_{n p q,h} \cM_{,mh}\cH_{n p r,k} R_{m r k q}\nonumber\\&&+2 \cH^T_{m n q,h} \cM_{,rk} \cH_{n p q,k} R_{p r h m}-2 \cH^T_{m n p,k} \cM_{,rm} \cH_{n p q,h} R_{q r h k}\big]\labell{F1F3}
\eeqa 
Each term is invariant under the $SL(2,{R})$ transformations. For zero axion background, each term has  the following couplings:
\beqa
\cH^T_{h q r,n}\cM_{,rm}\cH_{k n p,h}R_{m p k q}&=&2e^{\phi_0}F_{r,m}(F_{h q r,n}H_{k n p,h}+H_{h q r,n}F_{k n p,h})R_{m p k q}\labell{HMH}\\
&&+\sqrt{2}\kappa\phi_{,rm}(e^{\phi_0}F_{h q r,n}F_{k n p,h}-e^{-\phi_0}H_{h q r,n}H_{k n p,h})R_{m p k q}\nonumber
\eeqa
The couplings corresponding to the terms in the first line of \reef{HMH} have been found in \cite{Garousi:2013lja}. The couplings corresponding to the last term above have been found in \cite{Garousi:2013tca}. The couplings corresponding to the first term in the second line of \reef{HMH} are the couplings $F^{(3)}F^{(3)}R\phi$ in the Einstein frame.   We have checked it explicitly that they are   consistent with \reef{FnRD} for $n=3$.

\subsection{$F^{(n)}F^{(n)}\phi\phi$} 

Replacing the tensors $T_i$ for $ n=m $ case which are calculated in \reef{tn=m}, into \reef{kin3}, one finds the following result for the kinematic factor \reef{kin2} when both the NSNS polarization tensors are   the dilaton polarization \reef{dilpol}: 
\beqa
\cK &=& \frac{\phi_3\phi_4}{2^{11}} \bigg[(n-5) s t u F_{12}+4 n F_{12}^{\alpha \mu} \bigg((n-5) s u k_{1\mu } k_{4\alpha }\nn\\
&&\qquad\quad+(n-5) s t k_{2\alpha }k_{4\mu }+[(n-5)^2 s^2-8 t u] k_{4\alpha }k_{4\mu } \bigg)  \bigg]\labell{FnD}
\eeqa
We are going to find the couplings with structure $ F^{(n)} F^{(n)} \phi \phi $ for $ n=1,2,3,4,5 $ which are consistent with the above kinematic factor.  
For $n=5$, only the last term survives. The coupling in the Einstein frame is
\beqa
S &\supset& \frac{1}{6}\frac{\gamma}{\kappa^2} \int d^{10}x \sqrt{-G} e^{-3\phi_0/2} \big[ 
 F_{h}{}_{pqrs}{}_{,m} F_{kpqrs}{}_{,n} \Phi_{,hk}\Phi_{,mn}  \big]\labell{F5D1}
\eeqa
which can easily be extended to the S-duality invariant form. However, the couplings in the string frame which can be studied under the linear T-duality, are not so easy to read from the kinematic factor \reef{FnD}.  So we consider  all possible on-shell couplings  with structure   $ F^{(5)} F^{(5)} \phi\phi $ in the string frame with unknown coefficients, \ie
\beqa
S &\supset& \frac{1}{2}\frac{\gamma}{\kappa^2}\int d^{10}x \sqrt{-G} \big[C_{1} F_{nkpqr,s} F_{nkpqr,s} \Phi_{,hm} \Phi_{,hm}\nn\\
&&+C_{2} F_{nkpqr,s} F_{nkpqs,r} \Phi_{,hm} \Phi_{,hm}+C_{3} F_{mkpqr,s} F_{nkpqr,s} \Phi_{,hm} \Phi_{,hn}\nn\\
&&+C_{4} F_{mkpqr,s} F_{nkpqs,r} \Phi_{,hm} \Phi_{,hn}+C_{5} F_{kpqrs,n} F_{mkpqr,s} \Phi_{,hm} \Phi_{,hn}\nn\\
&&+C_{6} F_{kpqrs,m} F_{kpqrs,n} \Phi_{,hm} \Phi_{,hn}+C_{7} F_{hnpqr,s} F_{mkpqr,s} \Phi_{,hm} \Phi_{,nk}\nn\\
&&+C_{8} F_{hnpqr,s} F_{mkpqs,r} \Phi_{,hm} \Phi_{,nk}+C_{9} F_{hnpqr,s} F_{mpqrs,k} \Phi_{,hm} \Phi_{,nk}\nn\\
&&+C_{10} F_{hpqrs,n} F_{mpqrs,k} \Phi_{,hm} \Phi_{,nk}+C_{11} F_{hpqrs,m} F_{npqrs,k} \Phi_{,hm} \Phi_{,nk}\nn\\
&&+C_{12} F_{hpqrs,n} F_{kpqrs,m} \Phi_{,hm} \Phi_{,nk}\big]\labell{allcontractions}
\eeqa
and find the coefficients by imposing the   constraint that the couplings in the Einstein frame are given by the above equation.

To find the  coefficients, one has to consider the string frame couplings with structure  $ F^{(5)} F^{(5)} R R $ which have  been found in \cite{Garousi:2013lja}, and the string frame couplings with structure  $ F^{(5)} F^{(5)} R\phi $ which have been found in \reef{mF5RD}. Both of them produce couplings with structure  $ F^{(5)} F^{(5)} \phi\phi $ when transforming   them to the Einstein frame \reef{transf}. Transforming all couplings to the Einstein frame and constraining them to be identical with    the coupling \reef{F5D1}, one finds the following couplings in the string frame: 
\beqa
S &\supset& \frac{\gamma}{\kappa^2}\int d^{10}x \sqrt{-G} \big[- \frac{1}{30} F_{mnpqr}{}_{,s} (F_{mnpqr}{}_{,s} \Phi_{,hk} - 10 F_{k}{}_{npqr}{}_{,s} \Phi_{,h}{}_{m})\Phi_{,hk}\big]\labell{mF5D}
\eeqa
Plus the following terms which contains some of the unknown coefficients:
 
\beqa
S &\supset& \frac{1}{2}\frac{\gamma}{\kappa^2}\int d^{10}x \sqrt{-G} \big[ C_{5} F_{kpqrs,n} F_{mkpqr,s} \Phi_{,hm} \Phi_{,hn}\nn\\
&&-\frac{1}{5} C_{2} F_{nkpqr,s} F_{nkpqr,s} \Phi_{,hm} \Phi_{,hm}+\frac{1}{10} C_{11} F_{nkpqr,s} F_{nkpqr,s} \Phi_{,hm}\Phi_{,hm}\nn\\
&&+C_{2} F_{nkpqr,s} F_{nkpqs,r} \Phi_{,hm} \Phi_{,hm}+C_{6} F_{kpqrs,m} F_{kpqrs,n} \Phi_{,hm} \Phi_{,hn}\nn\\
 &&-C_{5} F_{mkpqr,s} F_{nkpqr,s} \Phi_{,hm} \Phi_{,hn}-5 C_{6} F_{mkpqr,s} F_{nkpqr,s} \Phi_{,hm} \Phi_{,hn}\nn\\
&&-2 C_{11} F_{mkpqr,s} F_{nkpqr,s} \Phi_{,hm} \Phi_{,hn}+4 C_{5} F_{mkpqr,s} F_{nkpqs,r} \Phi_{,hm} \Phi_{,hn}\nn\\
&&+20 C_{6} F_{mkpqr,s} F_{nkpqs,r} \Phi_{,hm} \Phi_{,hn}+C_{7} F_{mkpqr,s} F_{nkpqs,r} \Phi_{,hm} \Phi_{,hn}\nn\\
&&-\frac{1}{2} C_{9} F_{mkpqr,s} F_{nkpqs,r} \Phi_{,hm} \Phi_{,hn}+6 C_{11} F_{mkpqr,s} F_{nkpqs,r} \Phi_{,hm} \Phi_{,hn}\nn\\
&&-2 C_{12} F_{mkpqr,s} F_{nkpqs,r} \Phi_{,hm} \Phi_{,hn}+C_{12} F_{hpqrs,n} F_{kpqrs,m} \Phi_{,hm} \Phi_{,nk}\nn\\
&&+C_{7} F_{hnpqr,s} F_{mkpqr,s} \Phi_{,hm} \Phi_{,nk}+C_{8} F_{hnpqr,s} F_{mkpqs,r} \Phi_{,hm} \Phi_{,nk}\nn\\
&&+C_{9} F_{hnpqr,s} F_{mpqrs,k} \Phi_{,hm} \Phi_{,nk}-C_{11} F_{hpqrs,n} F_{mpqrs,k} \Phi_{,hm} \Phi_{,nk}\nn\\
&&-C_{12} F_{hpqrs,n} F_{mpqrs,k} \Phi_{,hm} \Phi_{,nk}+C_{11} F_{hpqrs,m} F_{npqrs,k} \Phi_{,hm} \Phi_{,nk}\big]\labell{all}
\eeqa
However, using the Bianchi identity and the on-shell relations, one finds that they are zero. To see this explicitly, consider for example the terms with coefficient $C_2$, \ie
\beqa
-\frac{1}{5} C_{2}
F_{nkpqr,s}F_{nkpqr,s} \Phi_{,hm}\Phi_{,hm}+C_{2}F_{nkpqr,s}F_{nkpqs,r}\Phi_{,hm}\Phi_{,hm}
\eeqa
To apply the Bianchi identity we write the RR field strength in terms of the RR potential. To impose the on-shell relations, we first transform the couplings to the momentum space and then impose the on-shell relations. One finds  
\beqa
-48 C_{2} (k_1.k_2)^2 {C}_{1hmnp} {C}_{2npqr} k_{1q} k_{1r} k_{2h} k_{2m}
\eeqa
which can easily be observed that it is zero using the totally antisymmetric property of the RR potential. Simile calculation shows that all other terms in \reef{all} vanishes.

We now apply the T-duality transformations on the    couplings \reef{mF5D} to find the  string frame  couplings with structure  $F^{(4)}F^{(4)}\phi\phi$ in the type IIA theory. To this end,  we use the dimensional reduction on the  couplings  \reef{mF5D}  and  find the couplings with structure $ F^{(5)}_yF^{(5)}_y \phi\phi$. Under the linear T-duality transformations,    they  transforms to the couplings with structure $ F^{(4)}F^{(4)}\phi\phi$ and some other terms involving $R_{yy}$ in which we are not interested. The couplings with structure  $ F^{(4)}F^{(4)}\phi\phi$ are
\beqa
S &\supset& \frac{\gamma}{\kappa^2}\int d^{10}x \sqrt{-G} \big[- \frac{1}{6} F_{mnpq}{}_{,r} (F_{mnpq}{}_{,r} \Phi_{,hk} - \
8 F_{k}{}_{npq}{}_{,r} \Phi_{,h}{}_{m})\Phi_{,hk}\big]\labell{mF4D}
\eeqa
The transformation of the above couplings, the couplings in \reef{mF4RD} and the couplings with structure $F^{(4)}F^{(4)}RR$  (see eq.(26) in \cite{Garousi:2013lja}), to the Einstein frame produces the following couplings:
\beqa
S &\supset& \frac{1}{2}\frac{\gamma}{\kappa^2}\int d^{10}x \sqrt{-G} e^{-\phi_0} \big[-\frac{1}{48} F_{nkpq,r} F_{nkpq,r} \Phi_{,hm} \Phi_{,hm}\labell{EF4F4DD}\\&&\qquad\qquad\qquad\qquad\qquad+\frac{1}{6} F_{mkpq,r} F_{nkpq,r} \Phi_{,hm} \Phi_{,hn}+\frac{4}{3} F_{hpqr,n} F_{mpqr,k} \Phi_{,hm} \Phi_{,nk}\big]\nn
\eeqa
which are fully consistent with the kinematic factor \reef{FnD} for $n=4$. In writing the above result, we have used the Bianchi identity and the on-shell relations to simplify the result.

To find the string frame couplings with structure $F^{(3)}F^{(3)}\phi\phi$   in the type IIB theory, one has to use the dimensional reduction on the couplings \reef{mF4D} and find the couplings with structure $F^{(4)}_yF^{(4)}_y \phi\phi$. Then under the linear T-duality transformations, they transform to the following couplings with structure $F^{(3)}F^{(3)} \phi\phi$:
\beqa
S &\supset& \frac{\gamma}{\kappa^2}\int d^{10}x \sqrt{-G} \big[- \frac{2}{3}F_{mnp}{}_{,q} (F_{mnp}{}_{,q} \Phi_{,hk} - 6 \
F_{k}{}_{np}{}_{,q} \Phi_{,h}{}_{m})\Phi_{,hk}\big]\labell{mF3D}
\eeqa
We have checked that the transformation of the above couplings, the couplings in \reef{mF3RD} and the couplings with structure $F^{(3)}F^{(3)}RR$  (see eq.(20) in \cite{Garousi:2013lja}), to the Einstein frame produces the couplings with structure $ F^{(3)}F^{(3)}\phi\phi$ which are consistent with the kinematic factor \reef{FnD} for $n=3$. We will study the S-duality of these couplings in the next section.

Applying   the T-duality transformations on the   couplings \reef{mF3D}, one finds the following couplings in   the type IIA theory in the string frame: 
\beqa
S &\supset& \frac{\gamma}{\kappa^2}\int d^{10}x \sqrt{-G} \big[- 2F_{mn}{}_{,q} (F_{mn}{}_{,q} \Phi_{,hk} - 4 \
F_{k}{}_{n}{}_{,q} \Phi_{,h}{}_{m})\Phi_{,hk}\big]\labell{mF2D}
\eeqa
The transformation of the above couplings, the couplings in \reef{mF2RD} and the couplings with structure $F^{(2)}F^{(2)}RR$  (see eq.(21) in \cite{Garousi:2013lja}), to the Einstein frame produces the following couplings:
\beqa
S &\supset& \frac{1}{2}\frac{\gamma}{\kappa^2}\int d^{10}x \sqrt{-G} \big[9 F_{hk,p} F_{hm,n} \Phi_{,kp} \Phi_{,mn}+10 F_{hm,n} F_{nk,p} \Phi_{,hk} \Phi_{,mp}\labell{EF2F2DD}\\&&\qquad\qquad\qquad\quad-9 F_{hm,n} F_{hn,k} \Phi_{,kp} \Phi_{,mp}-F_{hk,p} F_{hm,n} \Phi_{,mp} \Phi_{,nk}
\big]\nn
\eeqa
which are consistent with the kinematic factor \reef{FnD} for $n=2$.  

Finally, applying  the T-duality transformations on the   couplings \reef{mF2D}, one finds the following couplings in  the  type IIB theory in the string frame: 
\beqa
S &\supset& \frac{\gamma}{\kappa^2}\int d^{10}x \sqrt{-G} \big[- 4F_{m}{}_{,q} (F_{m}{}_{,q} \Phi_{,hk} - 2 \
F_{k}{}_{}{}_{,q} \Phi_{,h}{}_{m})\Phi_{,hk}\big]\labell{mF1D}
\eeqa
We have checked that the transformation of the above couplings  and the couplings with structure $F^{(1)}F^{(1)}RR$  (see eq.(22) in \cite{Garousi:2013lja}), to the Einstein frame produces couplings with structure $ F^{(1)}F^{(1)}\phi\phi$ which are consistent with the kinematic factor \reef{FnD} for $n=1$.

\subsubsection{Consistency with the S-duality}

The Einstein frame couplings  $F^{(n)}F^{(n)}\phi\phi$ for $n=1,3,5$ are in the type IIB theory, so   they should be consistent with the S-duality. For $n=5$, the coupling is given in \reef{F5D1}. The S-duality invariant extension of this coupling is 
\beqa
S &\supset& -\frac{1}{12}\frac{\gamma}{\kappa^2}\int d^{10}x \sqrt{-G} E_{3/2} \bigg( 
 F_{h}{}_{pqrs}{}_{,m} F_{kpqrs}{}_{,n} \Tr[\cM_{,hk}\cM^{-1}_{,mn}]  \bigg)\labell{F5D2}
\eeqa
where the $SL(2,{R})$ invariant combination of the dilaton and the RR scalar is
\beqa
-\frac{1}{4}\Tr[\cM_{,hk}\cM^{-1}_{,mn} ]&=&2e^{2\phi_0}F_{h,k}F_{m,n}+\kappa^2\phi_{,hk}\phi_{,mn}\labell{FF}
\eeqa
The second term corresponds to the coupling \reef{F5D1}. The first term is the S-duality prediction for  the couplings of two RR 4-forms and two RR scalars. 

To study  the S-duality of couplings for $n=3$ case,   consider the following S-duality invariant action that has been found in \cite{Garousi:2013lja}:
\beqa
S &\supset&\frac{\gamma}{\kappa^2} \int d^{10}x\sqrt{-G}E_{3/2}\bigg[\frac{1}{6} \Tr[\cM_{,nh}\cM^{-1}_{,nk} ] \cH^T_{m p r,h}\cM_0 \cH_{m p r,k}\labell{F1HS}\\&&-\frac{1}{2} \Tr[\cM_{,nh}\cM^{-1}_{,km} ] \cH^T_{m p r,k} \cM_0\cH_{n p r,h}-\frac{1}{2} \Tr[\cM_{,hm}\cM^{-1}_{,kn} ] \cH^T_{m p r,k} \cM_0\cH_{n p r,h}\bigg]\nonumber
\eeqa
where in the presence of zero axion background, the $SL(2,{R})$ invariant $\cH^T\cM_0\cH$ has the following terms:
\beqa
\cH^T\cM_0\cH&=&e^{-\phi_0}HH+e^{\phi_0}F^{(3)}F^{(3)}
\eeqa
Using \reef{FF} and the above expression, one finds the S-duality invariant action \reef{F1HS}  has four different terms. Terms with structure $HH\phi\phi$ which have been verified by the corresponding S-matrix element in \cite{Garousi:2013tca}, terms with structure $HHF^{(1)}F^{(1)}$ which have been verified by the corresponding S-matrix element in section 3.2,  terms with structure $F^{(3)}F^{(3)}\phi\phi$ and  terms with structure $F^{(3)}F^{(3)}F^{(1)}F^{(1)}$. We have checked explicitly that the couplings in \reef{F1HS}  with structure $F^{(3)}F^{(3)}\phi\phi$ are consistent with the kinematic factor \reef{FnD} for $n=3$. The couplings in \reef{F1HS} with structure $F^{(3)}F^{(3)}F^{(1)}F^{(1)}$ are the prediction of the S-duality for the couplings of two RR 2-forms and two RR scalars.

To study  the S-duality of couplings for $n=1$ case, consider the following S-duality invariant action that has been found in \cite{Garousi:2013tca}:
 \beqa
S \!\supset\!\frac{\gamma}{\kappa^2 } \int d^{10}x \sqrt{-G} E_{3/2}\bigg[ a\bigg(\frac{1}{4} \Tr[\cM_{,nm}\cM^{-1}_{,nm} ]\bigg)^2 +\frac{b}{16} \Tr[\cM_{,nm}\cM^{-1}_{,hk} ]\Tr[\cM_{,hk}\cM^{-1}_{,nm} ]\bigg]\labell{F1HS1}
\eeqa
where the constants $a,b$ satisfy the relation $a+b=1$. Using the expression \reef{FF}, one finds the above action has three different couplings. The couplings with structure $\phi\phi\phi\phi$ which have been verified by the S-matrix element of four dilatons in \cite{Garousi:2013tca} and the couplings with structures $F^{(1)}F^{(1)}\phi\phi$ and $F^{(1)}F^{(1)}F^{(1)}F^{(1)}$. We have found that the couplings with structure $F^{(1)}F^{(1)}\phi\phi$ are reproduced by the kinematic factor \reef{FnD} for $n=1$. This fixes the constants to be $a=-1$ and $b=2$. The couplings in \reef{F1HS1} with structure $F^{(1)}F^{(1)}F^{(1)}F^{(1)}$ are the prediction of the S-duality for the couplings of four RR scalars. These couplings have been confirmed in \cite{Garousi:2013tca} to be consistent with the linear T-duality of the couplings with structure $F^{(3)}F^{(3)}F^{(3)}F^{(3)}$.

\section{Discussion}

In this paper, we have examined in details the calculation of the S-matrix element of two RR and two NSNS states in the RNS formalism to find the corresponding couplings at order $\alpha'^3$. For the gravity and B-field couplings, we have found perfect agreement with the eight-derivative couplings that have been found in \cite{Garousi:2013lja}. For the dilaton couplings in the Einstein frame, we have found that the couplings are fully consistent with the S-dual multiplets that have been found in \cite{Garousi:2013lja}. We have also found the couplings with structure $F^{(3)}F^{(1)}H\phi$ which are singlet under the $SL(2,{R})$ transformation.

Unlike the four NSNS couplings which have no dilaton in the string frame, we have found that there are non-zero   couplings between the dilaton and the RR fields in the string frame. The couplings with structure $F^{(n)}F^{(n)}\phi\phi$    are the following:
\beqa
S &\!\!\!\supset\!\!\!& \frac{\gamma}{\kappa^2}\int d^{10}x \sqrt{-G} \bigg[\frac{8}{(n-1)!} F_{ka_1\cdots a_{n-1}}{}_{,s} F_{ma_1\cdots a_{n-1}}{}_{,s}\Phi_{,h}{}_{m}-\frac{4}{n!} F_{a_1\cdots a_n}{}_{,s} F_{a_1\cdots a_n}{}_{,s} \Phi_{,hk} \bigg]\Phi_{,hk} \nn
\eeqa
The couplings with structure $F^{(n)}F^{(n)}R\phi$    are the following:
\beqa
S &\supset& - \frac{8 }{(n-2)!} \frac{\gamma}{\kappa^2}\int d^{10}x \sqrt{-G} \bigg[  R_{h}{}_{mnp} \
F_{kt}{}_{a_1\cdots a_{n-2}}{}_{,n} F_{mpa_1\cdots a_{n-2}}{}_{,t}  \labell{mFnRD}\\ 
&& +  R_{h}{}_{mnp} F_{mta_1\cdots a_{n-2}}{}_{,p} F_{kn}{}_{a_1\cdots a_{n-2}}{}_{,t}+ (n-2) R_{h}{}_{mnp} F_{mpt a_1\cdots a_{n-3}}{}_{,s} \
F_{kns}{}_{a_1\cdots a_{n-3} }{}_{,t} \nn\\
&&- 3 R_{h}{}_{mnp} \
F_{mpa_1\cdots a_{n-2} }{}_{,t} F_{kn}{}_{a_1\cdots a_{n-2}}{}_{,t} + R_{h}{}_{m}{}_{k}{}_{n} F_{nta_1\cdots a_{n-2} }{}_{,s} \
F_{ms}{}_{a_1\cdots a_{n-2}}{}_{,t} \bigg]\Phi_{,hk}\nn
\eeqa
And the couplings with structure $F^{(n)}F^{(n-2)}H\phi$    are the following:
\beqa
S &\supset& -   \frac{8}{(n-3)!} \frac{\gamma}{\kappa^2}\int d^{10}x \sqrt{-G} \bigg[ (n-3) H_{h}{}_{rs}{}_{,m} F_{kqrs a_1\cdots a_{n-4}}{}_{,p} 
F_{mp a_1\cdots a_{n-4}}{}_{,q}  \nn \\ 
&&+ H_{h}{}_{rs}{}_{,m} F_{krs a_1\cdots a_{n-3}}{}_{,q} \
F_{m a_1\cdots a_{n-3}}{}_{,q}- \frac{1}{(n-2)} H_{h}{}_{rs}{}_{,k} F_{rsa_1\cdots a_{n-2}}{}_{,q} 
F_{a_1\cdots a_{n-2}}{}_{,q} \nn\\
&& - 2H_{hm}{}_{r}{}_{,s}F_{krs a_1\cdots a_{n-3}}{}_{,q} F_{ma_1\cdots a_{n-3} }{}_{,q}  + 2H_{hm}{}_{r}{}_{,s} F_{kqr a_1\cdots a_{n-3}}{}_{,s}  
F_{m a_1\cdots a_{n-3}}{}_{,q}\nn \\ 
&&\qquad\qquad\qquad\qquad\qquad\qquad+  
H_{k}{}_{qr}{}_{,s}F_{pqr a_1\cdots a_{n-3}}{}_{,s} F_{h}{}_{a_1\cdots a_{n-3} }{}_{,p} 
\bigg]\Phi_{,hk}\labell{mFnF3}
\eeqa
The number of indices $a_1\cdots a_{n-m}$ in the RR field strengths is such that the total number of the indices of $F^{(n)}$ must be $n$. For example, $F_{kqrs a_1\cdots a_{n-4}}{}_{,p}$ is  $F_{kqrs}{}_{,p}$ for $n=4$ and is zero for $n<4$. We have shown that the transformation of the above couplings and the couplings with structure $F^{(n)}F^{(n-2)}HR$ and $F^{(n)}F^{(n)}RR$  to the Einstein frame are fully consistent with the S-duality and with the corresponding S-matrix elements. The above couplings have been found in this paper  for $1\leq n \leq 5$. However, using the fact that  for $6\leq n \leq 9$, it is impossible to have couplings in which the RR field strengths have no contraction with each other, one finds that the consistency of the couplings with the linear T-duality requires the above couplings to be extended to  $1\leq n \leq 9$.

Using the pure spinor formalism, the S-matrix element of two NSNS and two RR states has been also calculated in \cite{Policastro:2006vt}. The kinematic factor in this amplitude is given by
\beqa
\sum_{M,N}u^{ijmnpqm'n'p'q'a_1\cdots a_{M}b_1\cdots b_N}\bar{R}_{mnm'n'}\bar{R}_{pqp'q'}F_{a_1\cdots a_M,i}F_{b_1\cdots b_N,j}
\eeqa
where $\bar{R}$ is the generalized Riemann curvature \reef{trans}, and the tensor $u$ is given in terms of the trace of the gamma matrices as
\beqa
u^{ijmnpqm'n'p'q'a_1\cdots a_{M}b_1\cdots b_N}&=&-32\frac{c_{M}c_{N}}{M!N!}\bigg[2\veps_Ng^{mp}g^{m'q'}g^{ip}g^{j(n'}g^{p')k}\Tr(\gamma^n\gamma^{a_1\cdots a_M}\gamma_k\gamma^{b_1\cdots b_{N}})\nn\\
&&\qquad\qquad -(\veps_M+ \veps_N)g^{m'q'}g^{iq}g^{j(n'}g^{p')k}\Tr(\gamma^{mnp}\gamma^{a_1\cdots a_M}\gamma_k\gamma^{b_1\cdots b_{N}})\nn\\
&&\qquad\qquad +\frac{1}{2}\veps_Ng^{i[q|}g^{jp'}\Tr(\gamma^{|mnp]}\gamma^{a_1\cdots a_M}\gamma^{m'n'p'}\gamma^{b_1\cdots b_{N}})\bigg]
\eeqa
where $c_p^2=(-1)^{p+1}/16\sqrt{2}$ and $\veps_N=(-1)^{\frac{1}{2}N(N-1)}$. Writing the above kinematic factor in terms of the independent variables, we have checked that it is exactly identical to the kinematic factor \reef{kin2} for the cases that $N=M$ and $N=M-4$ . However, for the case that $N=M-2$ the above result is different from \reef{kin2}. In fact the factor $(\veps_M+ \veps_N)$ in the second line above is zero for $N=M-2$ whereas the corresponding kinematic term in \reef{kin2}  which is $\cK_2+\cK_3$ is non-zero. We think there must be a typo in the above amplitude, \ie the factor $(\veps_M+ \veps_N)$ should be $(i^{N-M}\veps_M+\veps_N)$. With this modification, we find agreement with the kinematic factor \reef{kin2} even for $N=M-2$.

The S-duality invariant couplings in the sections 4.1.1 and 4.3.1 predict various couplings for four RR fields. These couplings may be confirmed by the details study of the S-matrix element of four RR vertex operators. This S-matrix element has been calculated in \cite{Policastro:2006vt} in the pure spinor formalism. This amplitude can also be calculated in the RNS formalism using the KLT prescription \reef{KLT} and  using the S-matrix element of four massless open string spinors which has been calculated in \cite{Schwarz:1982jn}. In both formalisms the amplitude involves various traces of the gamma matrices which have to be performed explicitly, and then one can compare the eight-derivative couplings with the four RR couplings predicted by the S-duality. We leave the details of this calculation  to the further works.

The consistency of the  NSNS couplings with the on-shell linear T-duality and S-duality has been used in \cite{Garousi:2013lja} and in the present  paper to find various four-field couplings involving the RR fields. Since the four-point function \reef{contact} has only contact terms, the Ward identities corresponding to the T-duality and the S-duality of the scattering amplitude appear as the  on-shell linear dualities is the four-field couplings. On the other hand, one may require the higher derivative couplings to be consistent with the nonlinear T-duality and S-duality without using the on-shell relations. This may be used to find the eight-derivative couplings involving more than four fields.   A step in  this direction has been taken in \cite{Garousi:2013qka} to find the gravity and the dilaton couplings which are consistent with the off-shell S-duality.   It would be interesting to extended the four-field on-shell couplings found in \cite{Garousi:2013lja} and in the present  paper to the couplings which are invariant under the off-shell T-duality and S-duality.

 {\bf Acknowledgments}:    This work is supported by Ferdowsi University of Mashhad under grant 3/27102-1392/02/25.   

\appendix

\section{Evaluation of traces for $n=m$ and $n=m+2$ }

In this appendix, we calculate the traces \reef{tr} for the cases that $n=m$ and $n=m+2$. 

For the   case that $n=m$, one can easily observes that $T_2=T_3$. We use the following algorithm for performing the trace $T_1$: There are three possibilities for contracting the  first gamma $\gamma^{\sigma}$   with the other gammas. It contracts with one of the gammas in $\gamma^{\mu_{1}\cdots\mu_{n}}$, with $\gamma^{\tau}$, and with one of the gammas in $\gamma^{\nu_{1}\cdots\nu_{n}}$. The symmetry factor and the sign of each contraction can easily be evaluated using the gamma algebra $\{\gamma^{\mu},\gamma^{\nu}\}=-2\eta^{\mu\nu}$. After performing these contractions, there is only one contraction for the leftover gammas. The leftover gammas for the first contraction is $\Tr(  \gamma^{\mu_{2}\cdots\mu_{n}}\gamma^{\tau} \gamma^{\nu_{1}\cdots\nu_{n}})$.  This trace can easily be evaluated because $\gamma^{\mu_{2}\cdots\mu_{n}}\gamma^{\tau}$ must be contracted with $\gamma^{\nu_{1}\cdots\nu_{n}}$. The reason is that the latter gammas are totally antisymmetric which can not contract among themselves. The leftover gammas for the second contraction is $\Tr(  \gamma^{\mu_{1}\cdots\mu_{n}} \gamma^{\nu_{1}\cdots\nu_{n}})$ which has again one contraction.  The leftover gammas for the third contraction is $\Tr(  \gamma^{\mu_{1}\cdots\mu_{n}}\gamma^{\tau} \gamma^{\nu_{2}\cdots\nu_{n}})$. There is only one possibility for the gammas $\gamma^{\tau} \gamma^{\nu_{2}\cdots\nu_{n}}$ to contract with the first bunch of gammas $\gamma^{\mu_{1}\cdots\mu_{n}}$. Taking the symmetry factors and the appropriate signs, one finds the   result for $T_1$. 

For the other traces, one should note that after performing the first contractions,  the leftover gammas have more than one contraction. However, the above algorithm can be used iteratively to reach to the trace which has only one contraction. After performing all the above contractions, one lefts with the trace of the identity matrix which is 32. The results   are    the following:  
\beqa
T_1^{\sigma \tau} &=& 16\frac{(-1)^{n(n+1)}a_{n}^{2}}{n!} \bigg[n \big(F_{12}^{\sigma \tau}+ F_{12}^{\tau \sigma}\big)-{\eta}^{\sigma \tau} F_{12} \bigg]\nn
\\
T_2^{\sigma \beta \rho \nu} &=& 16\frac{(-1)^{n(n+1)}a_{n}^{2}}{n!}\bigg[3 n {\eta}^{\sigma \beta} \big(F_{12}^{\rho \nu}- F_{12}^{\nu \rho}\big)\nn\\
&&+\frac{n!}{(n-2)!} \bigg(-F_{12}^{\nu \rho \beta \sigma}- F_{12}^{\rho \beta \nu \sigma}- F_{12}^{\rho \sigma \nu \beta}+ F_{12}^{\beta \sigma \nu \rho}+ F_{12}^{\nu \beta \rho \sigma}+ F_{12}^{\nu \sigma \rho \beta}\bigg)\bigg]\nn
\\
T_4^{\alpha \lambda \mu \beta \rho \nu} &=& 16\frac{(-1)^{n(n+1)}a_{n}^{2}}{n!} \bigg[-6 {\eta}^{\alpha \beta} {\eta}^{\lambda \rho} {\eta}^{\mu \nu} F_{12}+6n \bigg({\eta}^{\beta \lambda} {\eta}^{\mu \rho} F_{12}^{\nu \alpha}- {\eta}^{\alpha \beta} {\eta}^{\mu \rho} F_{12}^{\nu \lambda}\nn\\
&&+ {\eta}^{\alpha \beta} {\eta}^{\lambda \rho} F_{12}^{\nu \mu}+ {\eta}^{\beta \lambda} {\eta}^{\mu \rho} F_{12}^{\alpha \nu}- {\eta}^{\alpha \beta} {\eta}^{\mu \rho} F_{12}^{\lambda \nu}+ {\eta}^{\alpha \beta} {\eta}^{\lambda \rho} F_{12}^{\mu \nu}\bigg)+\frac{3n!}{(n-2)!} \nn\\
&& \times \bigg(-{\eta}^{\beta \mu} F_{12}^{\nu \rho \lambda \alpha}+ {\eta}^{\beta \lambda} F_{12}^{\nu \rho \mu \alpha}- {\eta}^{\alpha \beta} F_{12}^{\nu \rho \mu \lambda}+ {\eta}^{\beta \mu} F_{12}^{\rho \lambda \nu \alpha}- {\eta}^{\beta \lambda} F_{12}^{\rho \mu \nu \alpha}\nn\\
&&- {\eta}^{\beta \mu} F_{12}^{\rho \alpha \nu \lambda}+ {\eta}^{\alpha \beta} F_{12}^{\rho \mu \nu \lambda}+ {\eta}^{\beta \lambda} F_{12}^{\rho \alpha \nu \mu}- {\eta}^{\alpha \beta} F_{12}^{\rho \lambda \nu \mu}- {\eta}^{\beta \mu} F_{12}^{\lambda \alpha \nu \rho}\nn\\
&&+ {\eta}^{\beta \lambda} F_{12}^{\mu \alpha \nu \rho}- {\eta}^{\alpha \beta} F_{12}^{\mu \lambda \nu \rho}- {\eta}^{\beta \mu} F_{12}^{\nu \lambda \rho \alpha}+{\eta}^{\beta \lambda} F_{12}^{\nu \mu \rho \alpha}+ {\eta}^{\beta \mu} F_{12}^{\nu \alpha \rho \lambda}\nn\\
&&- {\eta}^{\alpha \beta} F_{12}^{\nu \mu \rho \lambda}- {\eta}^{\beta \lambda} F_{12}^{\nu \alpha \rho \mu}+ {\eta}^{\alpha \beta} F_{12}^{\nu \lambda \rho \mu}\bigg)+\frac{n!}{(n-3)!}  \bigg(F_{12}^{\nu \rho \mu \beta \lambda \alpha}- F_{12}^{\nu \rho \lambda \beta \mu \alpha}\nn\\
&&+ F_{12}^{\nu \rho \alpha \beta \mu \lambda}+ F_{12}^{\nu \rho \beta \mu \lambda \alpha}- F_{12}^{\rho \mu \lambda \nu \beta \alpha}+ F_{12}^{\rho \mu \alpha \nu \beta \lambda}- F_{12}^{\rho \lambda \alpha \nu \beta \mu}+ F_{12}^{\rho \beta \mu \nu \lambda \alpha}\nn\\
&&- F_{12}^{\rho \beta \lambda \nu \mu \alpha}+ F_{12}^{\rho \beta \alpha \nu \mu \lambda}+ F_{12}^{\beta \mu \lambda \nu \rho \alpha}+ F_{12}^{\mu \lambda \alpha \nu \rho \beta}- F_{12}^{\beta \mu \alpha \nu \rho \lambda}+ F_{12}^{\beta \lambda \alpha \nu \rho \mu}\nn\\
&&+ F_{12}^{\nu \mu \lambda \rho \beta \alpha}- F_{12}^{\nu \mu \alpha \rho \beta \lambda}+ F_{12}^{\nu \lambda \alpha \rho \beta \mu}- F_{12}^{\nu \beta \mu \rho \lambda \alpha}+ F_{12}^{\nu \beta \lambda \rho \mu \alpha}- F_{12}^{\nu \beta \alpha \rho \mu \lambda}\bigg)\bigg]\labell{tn=m}
\eeqa
where we have used the following notation in the above equations:
\beqa
F_{12} &\equiv& F_{1\mu_{1}\cdots\mu_{n}} F_2^{\mu_{1}\cdots\mu_{n}}\nn\\
F_{12}^{\sigma \tau}&\equiv& F_{1\mu_{1}\cdots\mu_{n-1}}{}^{ \sigma} F_2^{\mu_{1}\cdots\mu_{n-1} \tau}\nn\\
F_{12}^{\nu \rho \beta \sigma} &\equiv& F_{1\mu_{1}\cdots\mu_{n-2}}{}^{ \nu \rho} F_2^{\mu_{1}\cdots\mu_{n-2} \beta \sigma}\nn\\
F_{12}^{\nu \rho \mu \beta \lambda \alpha} &\equiv& F_{1\mu_{1}\cdots\mu_{n-3}}{}^{ \nu \rho \mu} F_2^{\mu_{1}\cdots\mu_{n-2} \beta \lambda \alpha}\labell{n=m}
\eeqa
Replacing the above tensors $T_1,\cdots, T_4$ into the kinematic factor \reef{kin2}, one finds the amplitude \reef{amp3} for two RR $n$-forms and two NSNS states. We have checked that the amplitude satisfies the Ward identity corresponding to the NSNS gauge transformations.   We have also checked that the kinematic factor vanishes  when one of the NSNS states is symmetric and the other one is antisymmetric.

For the case that $n=m+2$, one finds $T_2=-T_3$. Using the above algorithm, one finds the following results for the tensors $T_1,\, T_2,\, T_4$:
\beqa
T_1^{\sigma \tau} &=& 16\frac{(-1)^{n^2}a_{n}^2 }{(n-2)!}  F_{12}^{\tau \sigma }  \nn\\
T_2^{\sigma \beta \rho \nu} &=& 16\frac{(-1)^{n^2}a_{n}^2 }{(n-2)!}\bigg[3 {\eta}^{\beta \sigma}  F_{12}^{\nu  \rho }+(n-2)\big(  F_{12}^{\nu  \beta  \sigma \rho }-  F_{12}^{\nu  \rho  \sigma \beta }-  F_{12}^{\nu  \rho  \beta \sigma }-  F_{12}^{\rho  \beta  \sigma \nu }\big)\bigg]\nn\\
T_4^{\alpha \lambda\mu\beta \rho \nu} &=& 16\frac{(-1)^{n^2}a_{n}^2 }{(n-2)!}\bigg[6 \big({\eta}^{\lambda \beta } {\eta}^{\rho \mu }  F_{12}^{\nu  \alpha }- {\eta}^{\beta \alpha } {\eta}^{\rho \mu }  F_{12}^{\nu  \lambda }+ {\eta}^{\beta \alpha } {\eta}^{\rho \lambda }  F_{12}^{\nu  \mu }\big)+3 (n-2) \nn\\
&&\times \bigg(-  {\eta}^{\mu \beta }  F_{12}^{\nu  \lambda  \alpha \rho }+  {\eta}^{\lambda \beta }  F_{12}^{\nu  \mu  \alpha \rho }-  {\eta}^{\beta \alpha }  F_{12}^{\nu  \mu  \lambda \rho }-  {\eta}^{\mu \beta }  F_{12}^{\nu  \rho  \alpha \lambda }+  {\eta}^{\lambda \beta }  F_{12}^{\nu  \rho  \alpha \mu }\nn\\
&&+  {\eta}^{\mu \beta }  F_{12}^{\nu  \rho  \lambda \alpha }-  {\eta}^{\beta \alpha }  F_{12}^{\nu  \rho  \lambda \mu }-  {\eta}^{\lambda \beta }  F_{12}^{\nu  \rho  \mu \alpha }+  {\eta}^{\beta \alpha }  F_{12}^{\nu  \rho  \mu \lambda }+  {\eta}^{\mu \beta }  F_{12}^{\rho  \lambda  \alpha \nu }\nn\\
&&+  {\eta}^{\mu \beta }  F_{12}^{\nu  \rho  \lambda \alpha }-  {\eta}^{\beta \alpha }  F_{12}^{\nu  \rho  \lambda \mu }-  {\eta}^{\lambda \beta }  F_{12}^{\nu  \rho  \mu \alpha }+  {\eta}^{\beta \alpha }  F_{12}^{\nu  \rho  \mu \lambda }+  {\eta}^{\mu \beta }  F_{12}^{\rho  \lambda  \alpha \nu }\nn\\
&&-  {\eta}^{\lambda \beta }  F_{12}^{\rho  \mu  \alpha \nu }+  {\eta}^{\beta \alpha }  F_{12}^{\rho  \mu  \lambda \nu }\bigg)+\frac{(n-2)!}{(n-4)!}\bigg(  F_{12}^{\beta  \mu  \lambda  \alpha \nu  \rho }+  F_{12}^{\nu  \beta  \lambda  \alpha \rho  \mu }-  F_{12}^{\nu  \beta  \mu  \alpha \rho  \lambda }\nn\\
&&+  F_{12}^{\nu  \beta  \mu  \lambda \rho  \alpha }+  F_{12}^{\nu  \mu  \lambda  \alpha \rho  \beta }+  F_{12}^{\nu  \rho  \beta  \alpha \mu  \lambda }-  F_{12}^{\nu  \rho  \beta  \lambda \mu  \alpha }+  F_{\nu  \rho  \beta  \mu \lambda  \alpha }-  F_{\nu  \rho  \lambda  \alpha \beta  \mu }\nn\\
&&+  F_{\nu  \rho  \mu  \alpha \beta  \lambda }-  F_{12}^{\nu  \rho  \mu  \lambda \beta  \alpha }-  F_{12}^{\rho  \beta  \lambda  \alpha \nu  \mu }+  F_{12}^{\rho  \beta  \mu  \alpha \nu  \lambda }-  F_{12}^{\rho  \beta  \mu  \lambda \nu  \alpha }-  F_{12}^{\rho  \mu  \lambda  \alpha \nu \beta}\bigg)\bigg]\labell{tn=m+2}
\eeqa
where we have used among other things the fact that $a_{n-2}=a_{n}$. In above tensors, we have used the following notation for $F_{12}$s which are different from the ones in \reef{n=m}:
\beqa
F_{12}^{\sigma \tau}&\equiv& F_{1\mu_{1}\cdots\mu_{n-2}}{}^{ \sigma\tau} F_2^{\mu_{1}\cdots\mu_{n-2} }\nn\\
F_{12}^{\nu \rho \beta \sigma} &\equiv& F_{1\mu_{1}\cdots\mu_{n-3}}{}^{\nu \rho \beta} F_2^{\mu_{1}\cdots\mu_{n-3} \sigma}\nn\\
F_{12}^{\nu \rho \mu \beta \lambda \alpha} &\equiv& F_{1\mu_{1}\cdots\mu_{n-4}}{}^{\nu \rho \mu \beta} F_2^{\mu_{1}\cdots\mu_{n-4} \lambda \alpha}\labell{n=m+2}
\eeqa
Note that $F_1$ is $n$-form and $F_2$ is $(n-2)$-form. Replacing the above tensors $T_1,\cdots, T_4$ into the kinematic factor \reef{kin2}, one finds the amplitude \reef{amp3} for one RR $n$-form, one RR $(n-2)$-form  and two NSNS states. We have checked that the amplitude satisfies the Ward identity corresponding to the NSNS gauge transformations.    We have also checked that the kinematic factor vanishes  when one of the NSNS states is symmetric and the other one is antisymmetric.

\section{Massless poles  }

In this appendix we calculate the S-matrix element of two RR and two gravitons in the type II supegravities and compare the result with the leading order terms of the string theory S-matrix element \reef{amp3} which is given by the following amplitude:
\beqa
{\cal A}_{\rm low}=i\frac{2^5\kappa^2 e^{-2\phi_0}}{stu}\cK\labell{amplow}
\eeqa
where the explicit form of the kinematic factor $\cK$ for various external states has been given in sections 3 and 4.

The S-matrix element in field theory is independent of the field redefinitions which present  different forms of  the same theory. We use the democratic formulation of the supergravity which has been found in \cite{Fukuma:1999jt}. The RR part of this action is reproduced in the double field theory \cite{Hohm:2011zr}, so this part of the action is  manifestly invariant under the T-duality.  In the Einstein frame, the action is given as 
\beqa
S&=&\frac{1}{2\kappa^2} \int d^{10}x \sqrt{-g} \bigg(R-\frac{1}{2} \prt_{\mu} \Phi \prt^{\mu} \Phi-\frac{1}{2} e^{-\Phi} |H|^2-\frac{1}{4}\sum_{n=1}^{9} e^{\frac{1}{2}(5-n)\Phi}|\hat{F}_{n}|^2\bigg)\labell{sugra}
\eeqa
The nonlinear RR field strengths satisfy the following duality relation:
\beqa
\hat{F}_n=(-1)^{\frac{1}{2}(10-p)(9-p)}*\hat{F}_{10-n}
\eeqa
and are related to the RR potentials as
\beqa
\sum_{n=1}^{9} \hat{F}_{n}&=&e^{-B} \sum_{n=1}^{9} dC_{n-1}
\eeqa
where the product is the wedge product. The linear field strengths, \eg  $H=dB$, are defined such that 
\beqa
H_{\mu \nu \xi}
&=&\prt_{\mu} B_{\nu \xi} \pm\ {\rm cyclic\ permutations}
\nonumber
\eeqa
If A is a p-form, \ie $A=\frac{1}{p!}A_{\mu_1\cdots\mu_p}dx^{\mu_1}\wedge dx^{\mu_p}$,  and B is a q-form then
\beqa
(A \wedge B)_{\mu_1\cdots\mu_p \nu_1\cdots\nu_q} &=& \frac{(p+q)!}{p!q!}A_{[\mu_1\cdots\mu_p}B_{\nu_1\cdots\nu_q]}  
\eeqa
where the the bracket notation means antisymmetrization in the usual way, \eg $A_{[\mu}B_{\nu]}=\frac{1}{2}(A_{\mu}B_{\nu}-A_{\nu}B_{\mu})$. The square of the RR field strength in \reef{sugra} is $|\hat{F}_{n}|^2=\frac{1}{n!}\hat{F}_{\mu_1\cdots\mu_n}\hat{F}^{\mu_1\cdots\mu_n}$.

Given the supergravity \reef{sugra}, one can
calculate  different propagators, vertices, and subsequently the scattering amplitudes
for massless NSNS and RR states.
To have the normalizations of fields in \reef{sugra} to be consistent with the normalization of the string theory amplitude \reef{amp3}, we use the following   perturbations around the flat background, \ie 
\beqa
g_{\mu\nu}&=&\eta_{\mu\nu}+2\kappa h_{\mu\nu};\,B^{(2)}=2\kappa b^{(2)}\,;\,\,
\Phi\,=\,\phi_0+\sqrt{2}\kappa \phi\,;\,\,{C}^{(n)}\,=\,2\sqrt{2}\kappa c^{(n)}\labell{perturb}
\eeqa  
Note that the coefficient of the RR kinetic term in the conventional  form of the supergravity  is twice the RR kinetic term  of the action \reef{sugra}. On the other hand we have normalized the string theory amplitude \reef{amp3} to be consistent with the normalization of fields in \cite{Garousi:2013lja}. The normalization of fields in \cite{Garousi:2013lja} are consistent with the conventional  form of the supergravity. Therefore, we have added an extra factor of $\sqrt{2}$ in the normalization of  the RR potentials  in compare with the normalization of the RR fields in  \cite{Garousi:2013lja}. 

 The scattering amplitude of two NSNS and two RR fields in the supergravity is given by the following Feynman amplitude: 
\beqa
A=A_s+A_u+A_t+A_c\labell{ampf}
\eeqa
where the massless poles and contact term depends on the polarization of the external states. When the two NSNS states are dilaton, the massless poles in $s$- and $u$-channels, and the contact terms are given as
\beqa 
A_s&=&[\tV_{F_1^{(n)}F_2^{(n)}h}]^{\mu\nu}\,[\tG_h ]_{\mu\nu,\lambda\rho}\,[\tV_{h\phi_3\phi_4}]^{\lambda\rho}\nn\\
A_u&=&[\tV_{F_1^{(n)}\phi_3 C^{(n-1)}}]^{\mu_1\cdots\mu_{n-1}}\,[\tG_{C^{(n-1)}} ]_{\mu_1\cdots\mu_{n-1}}{}^{\nu_1\cdots\nu_{n-1}}\,[\tV_{C^{(n-1)} \phi_4 F_2^{(n)}}]_{\nu_1\cdots\nu_{n-1}}\nn\\
A_c&=&\tV_{F_1^{(n)}F_2^{(n)}\phi_3\phi_4}\labell{ampf1}
\eeqa
The massless pole in the $t$-channel is the same as $A_u$ in which the particle labels of the external dilatons are interchanged, \ie $A_t= A_u(3 \leftrightarrow 4)$. Using the supergavity \reef{sugra}, one finds the propagators and the vertices in the above amplitudes to be
\beqa
[\tG_h]_{\mu\nu,\lambda\rho}&=&
-\frac{i}{2k^2}\left(\eta_{\mu\lambda}\eta_{\nu\rho}+
\eta_{\mu\rho}\eta_{\nu\lambda}-\frac{1}{4}\eta_{\mu\nu}
\eta_{\lambda\rho}\right) \nn\\
\big[\tG_{C^{(n)}}\big]_{\mu_1\cdots\mu_n}{}^{\nu_1\cdots\nu_n}&=&-\frac{in!}{k^2}
\eta_{[\mu_1}{}^{\nu_1}\eta_{\mu_2}{}^{\nu_2}\cdots\eta_{\mu_n]}{}^{\nu_n}\nn\\
\big[\tV_{\phi_3 \phi_4 h}\big]^{\lambda \rho}&=&-2i\kappa \left(k_3^{(\lambda} k_4^{\rho)} - \frac{1}{2} k_3 \inn k_4 \eta^{\lambda \rho}\right)\nn\\
\big[\tV_{F_1^{(n)}F_2^{(n)}h}\big]^{\lambda\rho}&=& i\kappa\,
\frac{1}{n!}\left(2n\,F_{12}^{(\lambda\rho )} -
\eta^{\lambda\rho}\,F_{12} \right)\nn\\
\big[\tV_{F_1^{(n)}\phi_3 C^{(n-1)}}\big]_{\nu_1\cdots\nu_{n-1}}&=& i\kappa\frac{1}{\sqrt{2}(n-1)!}(5-n)\,F_{1\lambda\nu_1\cdots\nu_{n-1}}k^{\lambda}\nn\\
\tV_{F_1^{(n)}F_2^{(n)}\phi_3\phi_4}&=&-2i\kappa^2 \frac{1}{n!}\left(\frac{5-n}{2}\right)^2 F_{12}\labell{ver1}
\eeqa
where we have used the notation \reef{n=m}. Replacing them in \reef{ampf1}, one finds the following results: 
\beqa
A_s&=& -\frac{4i\kappa^2}{n!s}\bigg(4 n (F_{12})_{\lambda\rho} k_3^{\lambda}k_4^{\rho} -\frac{1}{2}(5n-9)F_{12} k_3\inn k_4\bigg) \nn\\
A_u&=&\frac{2i\kappa^2n}{ (n-1)!u}(5-n)^2(F_{12})_{\lambda\rho}(k_1+k_3)^{\lambda}(k_1+k_3)^{\rho }\nn\\
A_t&=&\frac{2i\kappa^2n}{ (n-1)!t}(5-n)^2(F_{12})_{\lambda\rho}(k_1+k_4)^{\lambda}(k_1+k_4)^{\rho }\nn\\
 A_c&=&-2i\kappa^2 \frac{1}{n!}\left(\frac{5-n}{2}\right)^2 F_{12} 
\eeqa
Replacing these result into  \reef{ampf}, one finds the  field theory scattering amplitude of two RR and two dilatons which should be compared with the corresponding amplitude in \reef{amplow}. When we use the on-shell relations to write both amplitudes in terms of the independent variables, we find exact agreement between the two amplitudes for $n=1,2,3,4,5$.

When the two NSNS states are B-fields, the massless poles in $s$- and $u$-channels, and the contact terms are given as
\beqa 
A_s&=&[\tV_{F_1^{(n)}F_2^{(n)}h}]^{\mu \nu}[\tG_{h} ]_{\mu \nu, \lambda\rho}[\tV_{hb_3b_4}]^{\lambda\rho}+[\tV_{F_1^{(n)}F_2^{(n)}\phi}][\tG_{\phi} ][\tV_{\phi b_3 b_4}]\nn\\
A_u&=&[\tV_{F_1^{(n)}b_3 C^{(n-3)}}]^{\mu_1\cdots\mu_{n-3}}[\tG_{C^{(n-3)}} ]_{\mu_1\cdots\mu_{n-3}}{}^{\nu_1\cdots\nu_{n-3}}[\tV_{C^{(n-3)} b_4 F_2^{(n)}}]_{\nu_1\cdots\nu_{n-3}}\nn\\&&+[\tV_{F_1^{(n)}b_3 C^{(n+1)}}]^{\mu_1\cdots\mu_{n+1}}[\tG_{C^{(n+1)}}]_{\mu_1\cdots\mu_{n+1}}{}^{\nu_1\cdots\nu_{n+1}}[\tV_{C^{(n+1)} b_4 F_2^{(n)}}]_{\nu_1\cdots\nu_{n+1}} \nn\\
A_c&=&\tV_{F_1^{(n)}F_2^{(n)}b_3b_4}\labell{ampf2}
\eeqa
The vertices in the first line of $A_u$ are non-zero for $n \geq 3$. The graviton and the RR propagators, as well as the vertex $[\tV_{F_1^{(n)}F_2^{(n)}h}]^{\mu \nu}$ have been given in \reef{ver1}. The dilaton propagator and all other vertices can be calculated from \reef{sugra}. They are
\beqa
\tG_\phi &=&-\frac{i}{k^2}\nn\\
(\tV_{b_3b_4h})^{\lambda\rho}&=&-2i\kappa\,\left[
\frac{1}{2}\left(k_3\inn
k_4\,\eta^{\lambda\rho}-k_3^{\lambda}\,k_4^{\rho}-k_3^{\rho}\,k_4^{\lambda}
\right)\Tr( b_3\inn b_4)\right.
\nonumber\\
&&\quad-k_3\inn b_4\inn b_3\inn k_4\, \eta^{\lambda\rho}+
2\,k_3^{(\lambda}\,{ b_4}^{\rho)}\inn b_3\inn k_4
+2\,{k_4}^{(\lambda}\,{ b_3}^{\rho)}\inn b_4\inn k_3
\nonumber\\
&&\quad\left.+2k_3\inn{ b_4}^{(\lambda}\,{ b_3}^{\rho)}\inn k_4
-k_3\inn k_4\,( b_3^{\lambda}\inn b_4^{\rho}+ b_4^{\lambda}\inn b_3^\rho)
\vphantom{\frac{1}{2}}\right]
\labell{hbb}\\
\tV_{b_3b_4\phi}&=&-i{\sqrt{2}\ka}\left[2k_3\inn b_4\inn b_3\inn k_4
-k_3\inn k_4\Tr( b_3\inn b_4)\right]\nn\\
\tV_{F_1^{(n)}F_2^{(n)}\phi}&=&
-i\kappa\frac{1}{\sqrt{2}n!}(5-n)\,F_1\inn F_2\nn\\
(\tV_{F_1^{(n)} b_3  C^{(n-3)}})_{\nu_1\cdots\nu_{n-3}}&=&-i\kappa\frac{1}{(n-3)!}F_{1\lambda\rho\mu\nu_1\cdots\nu_{n-3}}b_3^{\lambda\rho}k^{\mu}\nn\\
(\tV_{F_1^{(n)} b_3  C^{(n+1)}})_{\nu_1\cdots\nu_{n+1}}&=& -2i\kappa\frac{1}{(n+1)!}k^{\lambda} b_{3\lambda[\nu_1}F_{1\nu_2\cdots\nu_{n+1}]}\nn\\
\tV_{F_1^{(n)}F_2^{(n)}b_3b_4}&=&-\frac{2i\kappa^2(n+2)(n+1)}{n!}b_{3[\mu_1\mu_2}F_{1\mu_3\cdots\mu_{n+2}]}b_4^{\mu_1\mu_2}F_2^{\mu_3\cdots\mu_{n+2}}+(3\leftrightarrow 4)\nonumber
\eeqa 
Replacing them in \reef{ampf2}, one finds the massless poles and the contact terms of the scattering amplitude of two B-fields and two RR fields. We have compared the resulting amplitude \reef{ampf} with the corresponding string theory amplitude \reef{amplow} and find exact agreement for $n=1,2,3,4,5$ when we write both amplitudes in terms of the independent variables. We have done similar calculations for all other external states and find agreement with the string theory result.

\end{document}